\documentclass[%
preprint,
superscriptaddress,
 amsmath,amssymb,
prb,
]{revtex4-2}
\usepackage[table,xcdraw]{xcolor}
\usepackage{graphics}% Include figure files
\usepackage{dcolumn}% Align table columns on decimal point
\usepackage{bm}% bold math

\usepackage{array}
\usepackage{booktabs}
\usepackage{multirow}
\usepackage{esint}
\usepackage{color}
\usepackage{graphicx}
\usepackage{epstopdf}
\usepackage{verbatim}
\usepackage{float}
\usepackage{subfig}
\usepackage{fancyvrb}
\usepackage[export]{adjustbox}
\usepackage{svg} %used to add svg figures

\usepackage{tikz}
\usepackage{pgfplots}
\usepackage{pgfplotstable}
\pgfplotsset{compat = newest}
\usetikzlibrary{shapes.geometric, arrows}
\usepackage{hyperref}% add hypertext capabilities
\usepackage{gensymb}

\usepackage{scalerel}
\usepackage{tikz}
\usetikzlibrary{svg.path}
\definecolor{orcidlogocol}{HTML}{A6CE39}
\tikzset{
  orcidlogo/.pic={
    \fill[orcidlogocol] svg{M256,128c0,70.7-57.3,128-128,128C57.3,256,0,198.7,0,128C0,57.3,57.3,0,128,0C198.7,0,256,57.3,256,128z};
    \fill[white] svg{M86.3,186.2H70.9V79.1h15.4v48.4V186.2z}
                 svg{M108.9,79.1h41.6c39.6,0,57,28.3,57,53.6c0,27.5-21.5,53.6-56.8,53.6h-41.8V79.1z M124.3,172.4h24.5c34.9,0,42.9-26.5,42.9-39.7c0-21.5-13.7-39.7-43.7-39.7h-23.7V172.4z}
                 svg{M88.7,56.8c0,5.5-4.5,10.1-10.1,10.1c-5.6,0-10.1-4.6-10.1-10.1c0-5.6,4.5-10.1,10.1-10.1C84.2,46.7,88.7,51.3,88.7,56.8z};
  }
}
\newcommand\orcidicon[1]{\href{https://orcid.org/#1}{\mbox{\scalerel*{
\begin{tikzpicture}[yscale=-1,transform shape]
\pic{orcidlogo};
\end{tikzpicture}
}{|}}}} %Added by SP ends

\newcommand{\gz}{\mathbb{Z}}

\newcommand{\bfr}{{\bf r}}

\newcommand{\bfx}{{\bf x}}

\newcommand{\beq}{\begin{equation}}
\newcommand{\eeq}{\end{equation}}
\newcommand{\beqs}{\begin{eqnarray}}
\newcommand{\eeqs}{\end{eqnarray}}
\newcommand{\beql}{\begin{equation} \label}

\newcommand{\half}{\frac{1}{2}}

\newcommand{\calD}{{\cal D}}

\newcommand{\calF}{{\cal F}}
\newcommand{\calG}{{\cal G}}

\newcommand{\calP}{{\cal P}}

\let\oldFootnote\footnote
\newcommand\nextToken\relax

\renewcommand\footnote[1]{%
    \oldFootnote{#1}\futurelet\nextToken\isFootnote}

\newcommand\isFootnote{%
    \ifx\footnote\nextToken\textsuperscript{,}\fi}

\DeclareMathOperator{\arctantwo}{arctan2}

\begin{document}
\title{Machine learning based prediction of the electronic structure of quasi-one-dimensional materials under strain}
\author{Shashank Pathrudkar \orcidicon{0000-0001-8546-8056}}
\affiliation{Department of Mechanical Engineering--Engineering Mechanics, Michigan Technological University}
\author{Hsuan Ming Yu \orcidicon{0000-0002-1227-942X}}
\affiliation{Department of Materials Science and Engineering, University of California, Los Angeles, CA 90095, USA}
\author{Susanta Ghosh \orcidicon{0000-0002-6262-4121}}
\affiliation{Department of Mechanical Engineering--Engineering Mechanics, Michigan Technological University}
\affiliation{Faculty member of the Center for Data Sciences, Michigan Technological University}
% \affiliation{Faculty member of the Center for Applied Mathematics and Statistics, Michigan Technological University}
\author{Amartya S. Banerjee \orcidicon{0000-0001-5916-9167}}
\email{asbanerjee@ucla.edu}
\affiliation{Department of Materials Science and Engineering, University of California, Los Angeles, CA 90095, USA}
\date{\today}% It is always \today, today,

\begin{abstract}
We present a machine learning based model that can predict the electronic structure of quasi--one--dimensional materials while they are subjected to deformation modes such as torsion and extension/compression. The technique described here applies to important classes of materials  systems such as nanotubes, nanoribbons, nanowires, miscellaneous chiral structures and nano--assemblies, for all of which, tuning the interplay of mechanical deformations and electronic fields --- i.e., strain engineering --- is an active area of investigation in the literature. Our model incorporates global structural symmetries and atomic relaxation effects, benefits from the use of \textit{helical coordinates} to specify the electronic fields, and makes use of a specialized data generation process that solves the symmetry-adapted equations of Kohn-Sham Density Functional Theory in these coordinates. Using armchair single wall carbon nanotubes as a prototypical example, we demonstrate the use of the model to predict the fields associated with the ground state electron density and the nuclear pseudocharges, when three parameters --- namely, the radius of the nanotube, its axial stretch, and the twist per unit length --- are specified as inputs. Other electronic properties of interest, including the ground state electronic free energy, can be evaluated from these predicted fields with low-overhead post-processing, typically to chemical accuracy. {Additionally, we show how the nuclear coordinates can be reliably determined from the predicted pseudocharge field using a clustering based technique.} Remarkably, only about $120$ data points are found to be enough to predict the three dimensional electronic fields accurately, which we ascribe to the constraints imposed by symmetry in the problem setup, the use of low-discrepancy sequences for sampling, and efficient representation of the intrinsic low-dimensional features of the electronic fields. We comment on the interpretability of our machine learning model and anticipate that our framework will find utility in the automated discovery of low--dimensional materials, as well as the multi-scale modeling of such systems. 
\end{abstract}


\maketitle

\section{Introduction}
\label{sec:Introduction}
Over the last decade, machine learning (ML) models have percolated into all areas of science and engineering. Indeed, data-driven research is already an important part of the medical sciences \cite{ching2018opportunities, mamoshina2016applications, thiagarajan2021explanation}, chemistry \cite{prezhdo2020advancing, richardson2016clinical}, and engineering fields like manufacturing \cite{wuest2016machine, pham2005machine}, applied thermodynamics \cite{chen2012artificial, wang2018prediction}, and miscellaneous others (see e.g. \citep{raissi2019physics, liu2020machine, brunton2020machine, mattey2022novel}). The recent interest in these techniques has been driven by the improvement in the machine learning algorithms themselves, as well as an exponential growth in computation power, and the abundance of data. Additionally, data analysis tasks such as regression, classification and  dimensionality reduction, which are commonly used across all areas of science, are easily handled by machine learning algorithms by their innate design \cite{bishop2006pattern, efron2016computer}, and this has contributed to the wide applicability of machine learning techniques.

% Machine learning methods have been intensely investigated due to the improvement in algorithms, exponential growth in computation power, and abundance of data that can potentially accelerate calculations. 

Machine learning methods have also shown great promise for various materials physics problems \cite{rajan2015materials,  correa2018accelerating,liu2017materials,schmidt2019recent, butler2018machine, na2020tuplewise, wei2019machine, schleder2019dft}. In particular, the use of high-throughput Density Functional Theory (DFT) \citep{KohnSham_DFT, HK_DFT} calculations in conjunction with machine learning techniques, has attracted much attention as a powerful tool for materials discovery \citep{kim2018machine, ong2019accelerating, kim2016organized, li2019thermodynamic, de2017use}. A large section of the research in this direction so far, has been aimed at predicting specific material properties and screening novel materials for targeted applications such as energy storage. This includes electronic properties like the band gap, chemical properties like adsorption and formation energies, and mechanical properties like Young's modulus and fracture toughness  \cite{pilania2016machine, rajan2018machine,chibani2020machine,toyao2018toward,takahashi2018rapid,de2016statistical,evans2017predicting,calfa2016property, liu2020machine}. A common feature of most of these predicted material properties is that they are \textit{low--dimensional} --- usually, simple scalars. An alternative to these approaches is to use machine learning to directly predict electronic fields such as the ground state electron density for atomic configurations of interest. This is appealing since such fields contain all the information for predicting various material properties --- at least in principle, and the machine learning model provides a way to bypass expensive DFT calculations which can compute these fields. Recent work in this direction includes \citep{moreno2020deep, brockherde2017bypassing, grisafi2018transferable, chandrasekaran2019solving, kamal2020charge, bogojeski2018efficient, fabrizio2019electron, zepeda2021deep, tsubaki2020quantum, alred2018machine, gong2019predicting, shi2021machine, ellis2021accelerating}. The large majority of these contributions have focused on molecular systems (e.g. hydrocarbon chains and clusters), while a few have considered bulk materials. The current contribution can be viewed as an extension of the aforementioned efforts of machine learning based prediction of electronic fields to broader classes of nanomaterials --- specifically, quasi-1D nanostructures. Notably, a separate strand of work has also explored improving Density Functional Theory predictions themselves, by trying to learn the elusive Hohenberg-Kohn functional \citep{HK_DFT} or by improving exchange correlation functionals used in Kohn-Sham theory \citep{brockherde2017bypassing, nagai2018neural, kanungo2019exact, schmidt2019machine, kanungo2021comparison}. This latter class of developments will not have much bearing on the discussion that follows below.

Although machine learning based prediction of the electronic fields appears to be an attractive option for the aforementioned reasons, the high--dimensional nature of the fields usually makes it necessary to generate large amounts of data for model training and validation purposes. Additionally, since the models require a description of the atomic environments as input, it becomes necessary to choose a cutoff radius for limiting the size of the environments, or to focus on small sized systems, in order to make the models tractable. Furthermore, a careful choice of the atomic environment descriptors needs to be made to enforce symmetry and locality properties \citep{zepeda2021deep}. From this perspective, the current contribution is quite distinct in that global structural symmetries in lieu of environmental descriptors are utilized here, and strains are employed as model inputs. Our approach is related in spirit to  \citep{shi2021machine} where the authors investigated machine learning models for the electronic fields in a hexagonal close packed crystalline material.

We present here a machine learning model that can predict the electronic structure of quasi-one-dimensional materials as they are subjected to strains commensurate with their geometries. One of the key motivations of our work is that the complex interplay of electronic fields and mechanical deformations in low--dimensional materials is an active area of investigation in the literature \citep{dai2019strain, si2016strain, schlom2014elastic, pereira2009strain, li2014elastic, jiang2015strain, ghassemi2012field, jiang2015strain}, and therefore, it is desirable to have machine learning models where strain parameters can be mapped to electronic fields for such systems. Additionally, the techniques described here are likely to find use in the discovery of novel phases of low--dimensional chiral matter \citep{aiello2022chirality} and multiscale modeling \citep{hakobyan2012objective}. The data generation process for the ML model here is based on a recently formulated electronic structure calculation technique, that exploits the global symmetries of quasi-one-dimensional structures, and enables Kohn-Sham DFT calculations for such systems using a few representative atoms in a symmetry adapted unit cell \citep{My_PhD_Thesis, banerjee2021ab, banerjee2016cyclic, PhysRevB.100.125143, yu2022density, agarwal2021spectral, agarwal2021HelicES}. This computational method, called Helical Density Functional Theory (Helical DFT), solves the symmetry adapted Kohn-Sham equations in so-called helical coordinates to yield electronic fields of interest, and is able to accommodate deformation modes such as extension, compression and torsion, commonly associated with tubular or wire-like nanostructures. Atomic relaxation effects as a response to the applied strains are automatically included, by driving the Hellman-Feynman forces \citep{Martin_ES} to zero. In order to map strain parameters to resultant electronic fields, we utilize a two-step machine learning model, motivated by recently developed techniques used to predict the high--dimensional deformation fields of multi walled carbon nanotubes \citep{yadav2021interpretable}. Specifically, we use Principal Component Analysis (PCA) to perform dimensionality reduction of the electronic fields, and a neural network to learn in the reduced space. Using armchair single-wall carbon nanotubes as an example, we demonstrate that the ML  model accurately predicts the ground-state electron density and the nuclear pseudocharge fields when the radius of the nanotube, its axial stretch, and the twist per unit length are provided as inputs. We have also developed a novel technique based on clustering that allows us to determine the nuclear coordinates from the ML model predicted nuclear pseudocharge field,
and we demonstrate the superior performance of this method when compared to alternatives. Other quantities of interest, including ground state energies and symmetry-adapted band diagrams can be readily computed from the ML model predicted fields through low-overhead postprocessing steps. The strategy of predicting smoothly varying ground state fields such as the electron density, and obtaining energies from this field, instead of predicting the latter directly, appears to work better in practice \citep{brockherde2017bypassing, shi2021machine}. In a similar manner, computation of the electronic bands using a non-self-consistent calculation involving the machine-learning based Hamiltonian (i.e., diagonalization of a symmetry adapted Kohn-Sham Hamiltonian, with the effective potential arising from machine learning predicted fields) is more straightforward when compared to prediction of the band diagram directly, as a function of the inputs. This is due to the complexities in the structure of the latter \citep{damle2019variational, marzari1997maximally}, including e.g., the appearance of band crossings associated with insulator-metal transitions.

In our example, only about $160$ simulations were performed, out of which around $120$ are used for training purposes. Yet, ground state energies could be typically predicted to chemical accuracy (i.e., to better than $1.6$ milli-Hartree per atom, or $1$-kcal/mol), band gap predictions were generally accurate to $0.02$ eV, while the band gap location was predicted accurately every time. This suggests that the predictions of three-dimensional electronic fields themselves are rather accurate even with this limited {training data}, a fact also directly borne out by the low normalized root mean square errors in these quantities. The {high accuracy of the present ML model} is likely related to (i) the constraints imposed by symmetry in the problem setup, (ii) efficient exploration of the input space through quasi-random low-discrepancy sequences, and (iii) significant reduction in the dimensionality of the electronic fields. Indeed, only $7$ and $15$ principal component modes were found to be sufficient to capture most of the variations in the ground state electron density and the nuclear pseudocharge fields, respectively, which reinforces points (i) and (iii) above. We also observed that the electronic fields and post-processed quantities are accurately predicted for inputs whose values were not used during training, thus suggesting that our model can predict anywhere in the input space, even beyond the training data. Notably, the machine learning surrogate model is much cheaper computationally --- while the DFT calculation can take up to hundreds of CPU hours (in order to include atomic relaxation effects through ab initio geometry optimization), the machine learning model prediction can be done in a fraction of a second, and the subsequent post-processing steps (including prediction of band diagrams) can be typically performed in about $30$ to $40$ minutes of wall time.

The rest of the paper is organized as follows. We first explain the scheme of the ab initio simulations, which are used to obtain the training data for our machine learning model. Details regarding the system under consideration and the governing equations are presented in Section \ref{sec:Data_gen_DFT}. This is followed by an overview of our machine learning model. Specifically, details of the dimension reduction of the electronic fields, neural network based regression, and a new approach to predict atomic coordinates are explained in Section \ref{sec:ML}. Post-processing of machine learning predicted electronic fields to evaluate various energy components, band structures and atomic coordinates is explained in Section \ref{sec:postprocessing}. Next, we {validate the} machine learning {model and quantify its accuracy} in Section \ref{sec:Results}. We also comment on the model interpretability. We end with our conclusions and a discussion of future research directions.

\section{Methodology: First principles calculations}
\label{sec:Data_gen_DFT}
In this section, we describe the system setup, key aspects of the first principles simulation method (Helical DFT).
% , and the postprocessing steps used for computing quantities of interest from the machine learning model predicted fields. 
The atomic unit system with $m_e = 1, \hbar = 1, \frac{1}{4 \pi \epsilon_0} = 1$ will be used throughout, unless otherwise mentioned.

For the rest of the paper, $\textbf{e\textsubscript{X}}$, $\textbf{e\textsubscript{Y}}$, $\textbf{e\textsubscript{Z}}$ will denote the standard orthonormal basis of $\mathbb{R}^3$. Vectors in three dimensions will be denoted using lowercase boldface letters, while $3\times3$ matrices will be denoted using uppercase boldface. Cartesian, cylindrical and helical  coordinates will be denoted as $(x,y,z)$, $(r, \vartheta, z)$, and $(r,\theta_1,\theta_2)$, respectively, and the relation between these is:
\begin{align}
\nonumber
r &= \sqrt{x^2 + y^2}\,,\,\theta_{1} =\frac{z}{\tau}\,,\\
\theta_{2} &= \frac{1}{2\pi}\arctantwo\,(y, x)-\alpha \frac{z}{\tau} = \frac{\vartheta}{2\pi} - \alpha \frac{z}{\tau}\,.
\label{eq:helical_coordinates}
\end{align}
Here, $\alpha$ is related to the twist in the system as explained below. 
\subsection{System specification and global symmetries}
We begin by providing a description of the geometry of the quasi-one-dimensional systems under study, and the associated computational domains. 
As a prototypical system, we consider a nanostructure aligned and infinite in extent along \textbf{e\textsubscript{Z}} . Since the system of interest is quasi-one-dimensional, it is of limited extent in the \textbf{e\textsubscript{X}}-\textbf{e\textsubscript{Y}} plane. These conditions imply that the system can be embedded in a cylinder with axis \textbf{e\textsubscript{Z}} (or annular cylinder, if the system is tubular --- as considered here), of infinite height and finite radius, and this region of space will be referred to as the \emph{global simulation domain}. The structures considered in this work may be undeformed, or more generally, they may include axial deformation (i.e., stretch or compression) along \textbf{e\textsubscript{Z}}, and/or torsional deformation about the same axis. As pointed out in the literature, helical and cyclic symmetries can be used to describe such systems conveniently \citep{James_OS, banerjee2021ab, yu2022density, Dumitrica_James_OMD, Pekka_Twisted_GNR, Pekka_CNT_Bending, CNT_Dumitrica}. Thus if the atoms of the system have positions:
\begin{align}
\mathcal{S} = \{ \mathbf{p}_1, \mathbf{p}_2, \mathbf{p}_3,\dots:\mathbf{p}_{i} \in \mathbb{R}^3\}\,,
\label{Eqn:S_coord}
\end{align}
then we may identify a discrete group of isometries:
\begin{align}
\mathcal{G} = \Big\{ \Upsilon_{m,n}=\big({\textbf{R}}_{(2\pi m\alpha+n\Theta)}|m\tau \textbf{e\textsubscript{Z}}\big):m \in \mathbb{Z},n=0,1,\dots,\mathfrak{N} - 1\Big\}\,,
\label{Eqn:symm_opera}
\end{align}
and a finite collection of atoms (called \textit{simulated atoms} or \textit{representative atoms}) with coordinates:
\begin{align}
\mathcal{P} = \{ \mathbf{r}_1, \mathbf{r}_2, \mathbf{r}_3,\dots,\mathbf{r}_{M}:\mathbf{r}_{i} \in \mathbb{R}^3\}\,,
\label{Eqn:P_coord}
\end{align}
such that the structure can be described as the orbit of the group $\mathcal{G}$ on the set $\calP$, i.e.,
\begin{align}
\mathcal{S} = \underset{n=0,1,\dots,\mathfrak{N}-1}{\bigcup_{m \in \mathbb{Z}}}\!\bigcup^{M}_{i=1} {\textbf{R}}_{(2\pi m\alpha+n\Theta)}\mathbf{r}_{i}+m\tau\,\textbf{e\textsubscript{Z}}\,.
\label{Eqn:sim_atoms}
\end{align}
Here, $\Upsilon_{m,n}$ is an isometry operation (i.e., rigid body motion) consisting of the rotation matrix ${\textbf{R}}_{(2\pi m\alpha+n\Theta)}$ with axis \textbf{e\textsubscript{Z}} and translation vector $m\tau \textbf{e\textsubscript{Z}}$. It acts on an arbitrary point $\mathbf{x} \in \mathbb{R}^3$ by rotating it through the angle $2\pi m\alpha+n\Theta$ about the axis of the nanostructure, while simultaneously translating it by $m\tau$ about the same axis. The quantity $\mathfrak{N}$ is a natural number that captures any ($\mathfrak{N}$-fold) cyclic symmetries in the nanostructure, and the angle $\Theta=2\pi/\mathfrak{N}$. The parameter $\tau$ is related to the axial pitch of the structure and it can capture extensions and compressions about the axis, while the scalar $\alpha$ is related to the rate of applied or intrinsic twist in the structure, measured as $\beta = 2\pi\alpha/\tau$. For the structures considered here, we have $0\leq\alpha<1$, with $\alpha = 0$ representing untwisted structures. For undeformed armchair carbon nanotubes, the value of $\tau$ as suggested by the ``roll-up construction'' \citep{CNT_5, Dumitrica_James_OMD} is $\sqrt{3} a$, where $a$ is the interatomic bond-length of graphene (Figure \ref{fig:graphene_rolling}). Note that the numbers $m \in \gz$ and $n \in \{0,1,\ldots, \mathfrak{N}-1\}$ introduced above serve to label the group elements of $\calG$ (i.e., the isometries $\Upsilon_{m,n}$).

As pointed out in \citep{banerjee2021ab, yu2022density}, a key advantage of the above  formulation is that with the knowledge of the relevant symmetry group, any quasi-one-dimensional material can be represented efficiently by means of the representative atoms alone --- usually, just a few are adequate, and the behavior of the system under deformations (small or large) can be obtained by minimizing the system's free energy with respect to the coordinates of the representative atoms. The Helical Density Functional Theory (Helical DFT) technique described below, provides a computational framework for carrying out this procedure.
\subsection{Helical Density Functional Theory (Helical DFT)}
\label{subsec:Helical_DFT}
We use Helical Density Functional Theory (Helical DFT) \citep{banerjee2021ab, yu2022density} to compute the electronic fields associated with the (possibly deformed) quasi-one-dimensional nanostructures of interest in this work. To accommodate the global symmetries of the system under study, Helical DFT solves the symmetry adapted equations of Kohn-Sham DFT within a \emph{fundamental domain} (or \emph{symmetry adapted unit cell}) that encapsulates the representative atoms. In the context of this work, provided that the simulated atoms have radial coordinates between $R_{in}$ and $R_{out}$, a suitable fundamental domain is the following region (expressed in cylindrical coordinates):
\begin{align}
\mathcal{D} = \Big\{ (r,\vartheta,z) \in \mathbb{R}^3:R_{in} \leq r \leq R_{out}, \frac{2\pi\alpha z}{\tau} \leq \vartheta \leq \frac{2\pi\alpha z}{\tau}+\Theta,0\leq z \leq \tau \Big\}\,.
\label{Eqn:funda_domain}
\end{align}

Due to the global symmetries of the system described above, the eigenstates of the Kohn-Sham Hamiltonian, and other quantities related to its spectrum can be labeled using the characters of the group (i.e., its complex one dimensional irreducible representations). For  $m \in \mathbb{Z}$ and $n  \in \{0,1,2,...,\mathfrak{N}-1\}$, these are \citep{My_PhD_Thesis, McWeeny,Hammermesh, banerjee2021ab}:
\begin{align}
\widehat{\mathcal{G}} = \Big\{ e^{2\pi i \big(m\eta+\frac{n \nu}{\mathfrak{N}} \big)}:\eta \in \left[-\frac{1}{2},\frac{1}{2}\right);\nu= \{0,1,\dots,\mathfrak{N} - 1 \}\Big\}\,.
\label{Eqn:g_widehat}
\end{align}
Accordingly, Helical DFT uses $(\eta,\nu)$ to label the eigenvalues, the eigenvectors, and the electronic occupations. For $j \in \mathbb{N}$, the symmetry adapted Kohn-Sham equations over the fundamental domain (i.e. $\bfx \in \calD$) are:
\begin{align}
\mathfrak{H}^{\text{KS}}\,\psi_j(\bfx;\eta,\nu) = \lambda_j(\eta,\nu)\,\psi_j(\bfx;\eta,\nu)\,,
\mathfrak{H}^{\text{KS}} = -\frac{1}{2}\Delta+V_{\text{xc}}+\Phi+\mathcal{V}_{\text{nl}}\,,
\label{Eqn:ks_eqn}
\end{align}
with the eigenstates $\psi_j(\bfx;\eta,\nu)$ satisfying the Helical Bloch conditions:
\begin{align}
\psi_j(\Upsilon_{m,n} \circ \bfx;\eta,\nu) = e^{-2\pi i \big({m \eta + \frac{n\nu}{ \mathfrak{N}} }\big)} \psi_j(\bfx;\eta,\nu)\,.    
\end{align}
In the above, $\mathfrak{H}^{\text{KS}}$ denotes the Kohn-Sham operator, $V_{xc}$ denotes the exchange correlation potential, $\Phi$ denotes the net electrostatic potential arising from the electrons and the nuclear pseudocharges (i.e., a combination of the Hartree and electron-nucleus interaction terms), and $\mathcal{V}_{nl}$ denotes the non-local pseudopotential operator. The field $\Phi$ obeys the following Poisson problem in terms of the electron density $\rho$ and the nuclear pseudocharge field $b$:
\begin{align}
-\Delta\Phi=4\pi\big(\rho+b\big)\,.
\label{Eqn:poisson}
\end{align}
The non-local pseudopotential operator can be expressed in Kleinman-Bylander form \cite{kleinman1982efficacious} as:
\begin{align}
\mathcal{V}_{\text{nl}}= \sum_{i=1}^{M}\sum_{p\in{\Gamma}_i}\gamma_{i,p}\hat{\chi}_{i,p}(\cdot;\eta,\nu;\mathbf{r}_i)\;\overline{\hat{\chi}_{i,p}(\cdot;\eta,\nu;\mathbf{r}_i)}\,,
\label{Eqn:nl_pp}
\end{align}
with $\chi_{i,p}$, $\gamma_{i,p}$ and ${\Gamma}_i$ denoting the atom-centered projection functions (associated with the $i^{th}$ atom), the corresponding normalization constants, and the total set of projectors for the atom,  respectively. Within Helical DFT, for a given set of atoms in the fundamental domain, the nuclear pseudocharge field $b$ and the non-local pseudopotential operator $\mathcal{V}_{\text{nl}}$ are computed explicitly, along with a suitable starting guess for the electron density $\rho$. Following these computations, the symmetry adapted Kohn-Sham equations (eq.~\ref{Eqn:ks_eqn}) are solved self-consistently \citep{banerjee2016periodic}. At self-consistency, the free energy per unit fundamental domain and the Hellman-Feynman forces on the atoms may be computed following the expressions presented in \citep{banerjee2021ab, yu2022density}.

It is important to point out at this stage that construction of the symmetry adapted Kohn-Sham Hamiltonian requires knowledge of the atomic coordinates within the fundamental domain, due to the explicit dependence of the operator $\mathcal{V}_{\text{nl}}$ on the latter. Thus, unless all-electron calculations are being performed, it is not possible to compute the Kohn-Sham eigenstates via a simple diagonalization step, even if the electron density and the nuclear pseudocharge fields are known. As discussed later (Section \ref{subsec:DBSCAN}), we address this issue in this work by means of an unsupervised learning technique that can pick out the atomic coordinates from the nuclear pseudocharge field, which in turn can be used to set up the operator $\mathcal{V}_{\text{nl}}$.
\subsection{Use of helical coordinates}
\label{subsec:helical_coordinates}
The fundamental domain $\calD$ assumes the form of a cuboid $[R_{in}, R_{out}] \times [0, 1] \times [0, 1/\mathfrak{N}]$ in helical coordinates. Helical DFT uses a higher order finite difference scheme in these coordinates to discretize and solve the governing equations \citep{banerjee2021ab, yu2022density}. Thus, the electronic fields computed by the method are available over a set of grid points (corresponding to the finite difference mesh) in the fundamental domain. 

In addition to converting the complicated geometry of the fundamental domain to a simple cuboidal geometry for simulations, helical coordinates allow for additional simplifications in the data generation process. First, irrespective of the nanotube radius and the level of torsional and axial deformation imposed, the helical coordinates of an atom within the fundamental domain are such that $\theta_1$ and $\mathfrak{N}\,\theta_2$ remain constant, as long as relaxation effects are negligible. Thus, even when relaxation effects are not small, this property can be used to provide good starting guesses to the structural relaxation procedure. Second, for nanotubes of any radii undergoing relatively small torsional or extensional distortions, the total number of grid points (and hence the size of the vector used for describing the electronic fields) can be kept constant, with relatively small changes to the overall accuracy of the calculations. To see this, we denote $\mathsf{N}_r, \mathsf{N}_{\theta_1}, \mathsf{N}_{\theta_2}$ as the number of grid points along the $r$, $\theta_1$ and $\theta_2$ directions, respectively. The electronic fields are then represented as vectors in dimension $\mathsf{N}_r \times \mathsf{N}_{\theta_1} \times \mathsf{N}_{\theta_2}$, and the mesh spacings corresponding to these discretization choices are:
\begin{align}
h_{r} = \frac{R_{out} - R_{in}}{\mathsf{N}_r}, h_{\theta_1} = \frac{1}{\mathsf{N}_{\theta_1}}, h_{\theta_2} = \frac{1/\mathfrak{N}}{\mathsf{N}_{\theta_2}}
\label{Eqn:mesh_spacing}
\end{align}
The overall mesh spacing $h = \text{max}\Big(h_{r},\tau h_{\theta_1},2\pi (\frac{R_{\text{in}}+R_{\text{out}}}{2})h_{\theta_2} \Big)$ dictates the accuracy of the calculation. In the radial direction, by enforcing a constant amount of vacuum padding around the tubes, the mesh spacing $h_{r}$ (and hence $\mathsf{N}_r$) can be kept constant with respect to the tube diameter. In the axial direction, small changes to $\tau$ with respect to its equilibrium value (due to imposed strains) do not affect the overall calculation accuracy appreciably, as long as $\mathsf{N}_{\theta_1}$ is large enough to accommodate the largest value of $\tau$ considered. Finally, in the $\theta_2$ direction, assuming the nanotube is placed halfway between $R_{\text{in}}$ and $R_{\text{out}}$, the effect of change in $\frac{R_{\text{in}}+R_{\text{out}}}{2}$ is offset by the corresponding change in cyclic group order $\mathfrak{N}$, thus helping keep the product $(\frac{R_{\text{in}}+R_{\text{out}}}{2})h_{\theta_2}$ constant. Thus the same value of $\mathsf{N}_{\theta_2}$ can be chosen irrespective of the tube diameter.
\subsection{Other details of first principles calculations}
\begin{figure}[htbp]
    \centering
    \includegraphics[trim={1cm 1cm 3cm 0.5cm}, clip, width=0.6\linewidth]{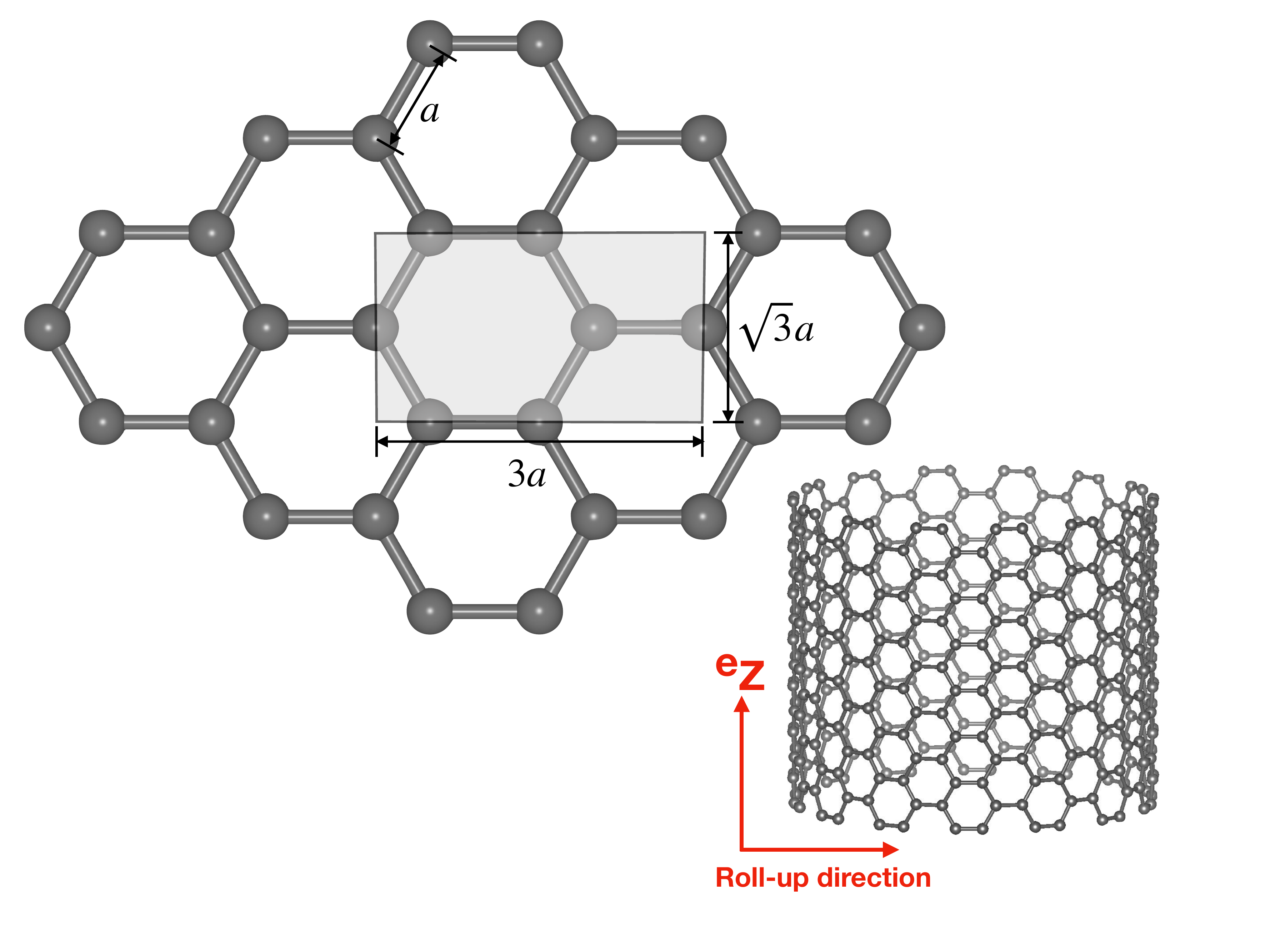}
    \caption{Roll-up construction of an undeformed armchair carbon nanotube, starting from a graphene sheet. The $4$ atoms shown in the shaded region are used for the data generation process using Helical DFT. The parameter \emph{a} represents the planar interatomic distance of $1.407$ Angstrom.}
    \label{fig:graphene_rolling}
\end{figure}

All Helical DFT calculations described in this work use a $4$-atom fundamental domain as shown in Figure \ref{fig:graphene_rolling}. To enable expeditious generation of data, calculations are done in two steps. First, ab initio geometry optimization calculations are done for a given level of axial and torsional strains by using $h = 0.3$ Bohr, and by sampling $15$ k-points in the $\eta$ direction. These discretization choices are sufficient to produce chemically accurate forces and ground state energies for the Troullier-Martins norm conserving pseudopotential \cite{troullier1991efficient} used to model the carbon atoms in this work \citep{yu2022density}. Atomic relaxation is carried out using the Fast Intertial Relaxation Engine \citep{bitzek2006structural}, and the structures are relaxed till each atomic force component drops below $0.001$ Ha/Bohr. Next, for each relaxed structure, we redo a self-consistent calculation to generate the electronic fields data for the machine learning model, using the finest discretization parameters that could be reliably afforded within computational resource constraints. This corresponds to a mesh spacing of $h = 0.25$ Bohr (resulting in $\mathsf{N}_r \times \mathsf{N}_{\theta_1} \times \mathsf{N}_{\theta_2} \approx 60,\!000$) and $21$ k-points in the $\eta$-direction. Due to the use of the above two-step procedure to generate the data, the machine learning model automatically incorporates atomic relaxation effects in response to applied strains.

For all ab initio calculations, we used the Perdew-Wang parametrization \cite{perdew1992accurate} of the Local Density Approximation \citep{KohnSham_DFT}, a $12^{\text{th}}$ order finite difference discretization scheme \citep{chelikowsky1994finite, Chelikowsky_Saad_1,  Chelikowsky_Saad_2, kikuji2005first, ghosh2017sparc_1, ghosh2017sparc_2, ghosh2019symmetry, banerjee2021ab}, vacuum padding of $11$ Bohrs in the radial direction and $1$ milli-Hartree of smearing using the Fermi-Dirac distribution.
\section{Methodology: Machine Learning Model for Prediction of the Electronic Fields} \label{sec:ML}
This section describes the proposed Machine Learning (ML) model that aims to predict the electronic structure (high--dimensional) of quasi--one--dimensional materials under torsional and axial loads. The tubular structures considered in this work can be characterized by their radius --- which is related to the degree of cyclic symmetry present in the structure, and the position of the atoms within the fundamental domain. Given strain parameters related to axial and torsional loads that the structure might be subject to, these atomic positions can be determined by minimizing the system's energy with respect to them. Thus, the trio of parameters $R_{\mathrm{avg}}$ --- the nanotube radius (or equivalently, the average radial coordinate of the atoms in the fundamental domain), $\alpha$ --- the twist parameter, and $\tau$ --- the axial pitch parameter, serve to specify a particular nanotube, along with the imposed torsional and axial strains. Accordingly, we let $\mathcal{H}$ denote the  map from the space consisting of system and loading parameters $(R_{\text{avg}}, \alpha, \tau)$ to the  electronic fields $\rho,b$ of the deformed nanotubes:
\begin{equation}
    \mathcal{H} : \{R_{\mathrm{avg}},\alpha,\tau\} \rightarrow \{\rho,b\} 
\end{equation} 
The objective of this work is to approximate this map $\mathcal{H}$ using a machine learning model. Inputs of this map $R_{\mathrm{avg}},\alpha$ and $\tau$ are scalars, while the outputs $\rho\,(r,\,\theta_1,\,\theta_2)$ and $b\,(r,\,\theta_1,\,\theta_2)$ are high--dimensional discretized scalar fields (expressed in helical coordinates). 

Approximating the map $\mathcal{H}$ directly through a supervised machine learning algorithm (such as  a Neural Network (NN)) is infeasible since the output quantities $\rho\,(r,\,\theta_1,\,\theta_2)$ and $b\,(r,\,\theta_1,\,\theta_2)$ are very high--dimensional. For instance, with the discretization choices adopted in this work, the field $\rho\,(r,\,\theta_1,\,\theta_2)$ is represented by a  vector of dimension close to $60,\!000$ (see Section \ref{sec:Data_gen_DFT}). 
% The difficulty in predicting such high--dimensional outputs using machine learning models is often referred to as as the \emph{curse of dimensionality} \cite{bishop2006pattern}. 
The difficulty in predicting such high--dimensional outputs using machine learning models is referred to as \emph{curse of dimensionality}. Specifically, the number of discrete cells required to discretize the output space grows exponentially with its dimensionality, and an exponentially large quantity of training data is then needed to ensure that the cells in the output space are accurately mapped from the input space \cite{bishop2006pattern}.

\begin{figure}[htbp]
    \centering
    \includegraphics[width=0.985\linewidth]{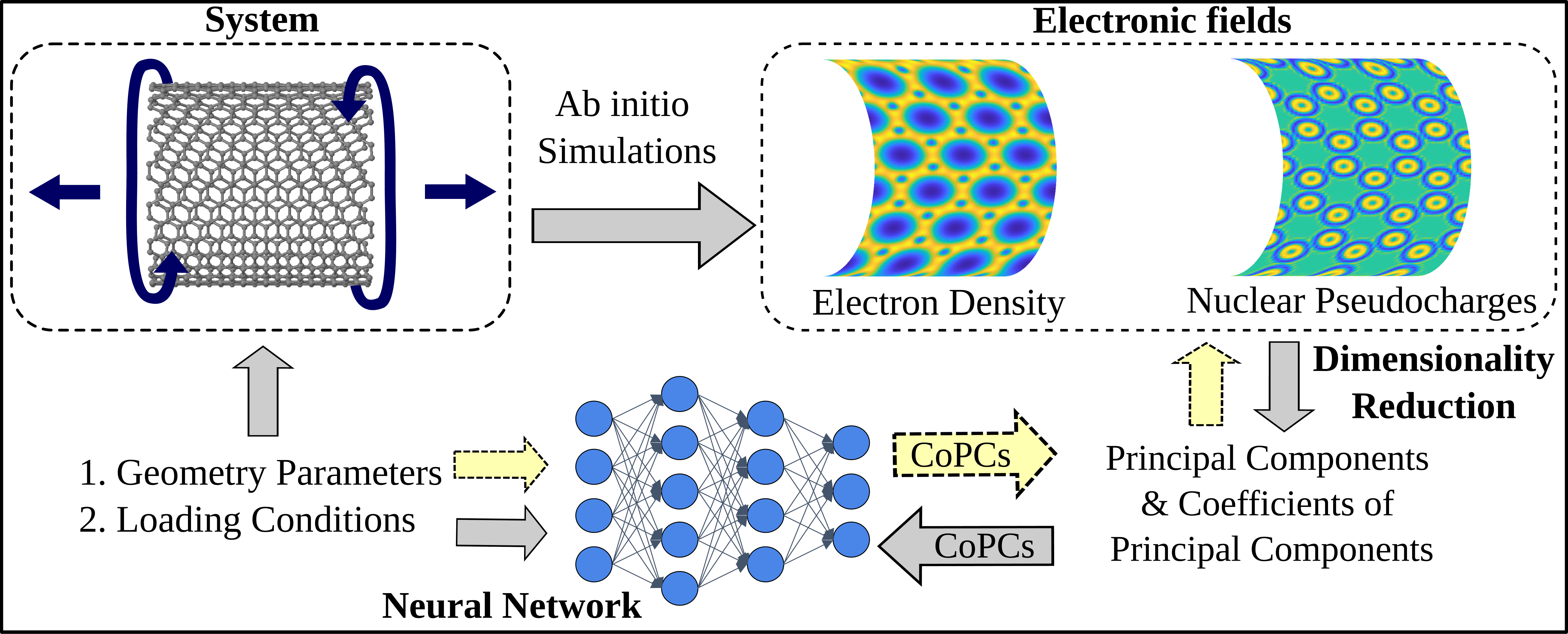}
    \caption{Schematic of the present Machine Learning (ML) model and the data generation process via DFT simulations. The firm arrows show the steps for data generation and training, and the dashed arrows show the steps for prediction via the ML model.}
    \label{fig:model_schematic}
\end{figure}

In the present work, we circumvent this problem by using Principal Component Analysis (PCA) to reduce the dimensions of the electronic fields. Subsequently, the low--dimensional representation of the electronic fields is learned via neural networks in a supervised manner.
This two-step approach, i.e., dimensionality reduction followed by learning in the reduced space, allows the prediction of the high--dimensional quantities such as electronic fields while reducing the data required for training. Schematic of the two-step ML model introduced above is given in Fig. \ref{fig:model_schematic}. Recently, a similar approach has been found to have excellent accuracy in high--dimensional predictions related to purely mechanical problems \cite{yadav2021interpretable, shapa2021deformation}. 

In the following sections, we detail various important aspects of the above ML model and also describe an auxiliary clustering based technique that allows us to {determine} the nuclear coordinates from the ML model predicted nuclear pseudocharge field. 
%This allows a lower--dimensional representation of data to capture the variation in the entire data through a small number of variables. 
% The PCA also provides a mapping from the high- to the low-dimensional representation and vice-versa. % as shown in the schematic for this model (Fig. \ref{fig:model_schematic}).
% \cite{yadav2021interpretable, bhattacharya2021model, shi2021machine, shapa2021deformation}
%The rest of this section describes the following constituents of the present ML model. Section \ref{subsec:Data_gen_ML} describes the exploration of the design space using the design of experiments. Section \ref{subsec:dim_red_pca} explains the dimensionality reduction technique used for the electronic fields. Section \ref{subsec:DNN} elaborates the supervised learning of low--dimensional representation of th electronic fields. Additionally, in section \ref{subsec:DBSCAN} we propose an unsupervised learning approach to predict nuclear coordinates 
\subsection{Design of Experiments to Explore the Input Space}
\label{subsec:Data_gen_ML}
We now describe the use of Design of Experiments (DoE) \cite{robbins1952some, lohr2019sampling} techniques for efficient sampling in the input space. As described above, the triplet of input parameters $\{R_{\mathrm{avg}},\alpha,\tau\}$ specify a particular nanotube and the applied strains. The number of possible combinations with these three input variables can be quite large even if finite bounds are emplaced for these variables. Given the relatively high cost of DFT simulations for the {deformed} nanotubes, it is infeasible to simulate nearly all possible combinations in the input space. Purely random sampling of the input space is not desirable either, since it may require a large number of sampling points to learn the pattern in the data  accurately \cite{fang2005design, santiago2012construction,simpson2001metamodels}.  To address this challenge, we generate sequences of quasi-random sampling points in the input space to reduce the number of simulations required for training an accurate ML model. 
%Furthermore, random sampling of the input space can curb the learning ability of the machine learning model or may require numerous additional data points to learn the pattern accurately \cite{fang2005design, santiago2012construction,simpson2001metamodels}. 
% To train our proposed ML model, we generate data through DFT simulations performed using the scheme given in Section \ref{sec:Data_gen_DFT}. Simulation data was obtained for 204 data points in the input variables space such that $R_{\mathrm{avg}} \in [20 \, \mathrm{Bohr}, 102 \, \mathrm{Bohr}] $, $\alpha \in [0,3.69\degree/\mathrm{nm}]$ and $\tau \in [4.4052,\; 4.8052]$.  The number of possible combinations with these three input variables is quite large.  Given the high cost of First Principles simulations, it is infeasible to simulate nearly all possible combinations to train the proposed ML model. To address this challenge and limit the number of simulations, we generate sequences of quasi–random sampling points in the input space. These sampling points guide the location in the input space where DFT simulations are to be performed. 
% Our input space consist of three design variables, and the bounds for these variables considered in this work are given above. We first formed a grid with 13, 6, and 5 grid points along $R_{\mathrm{avg}}$, $\alpha$, and $\tau$ direction respectively, which gave us a total of 390 grid points.  

\underline{Quasi-random sampling:} Space-filling designs can be used to explore the input domain effectively since they sample the space uniformly without assuming any prior knowledge of the problem \cite{lin2001sampling,fang2005design}. Commonly used space-filling designs include low discrepancy sequences \cite{hammersley1960monte,sobol1967distribution}, good lattice points \cite{mckay2000comparison}, Latin Hypercube Sampling \cite{jin2003efficient} and Orthogonal Latin Hypercube sampling \cite{owen1992orthogonal}. These methods are often evaluated based on their measure of uniformity \citep{santiago2012construction, johnson1990minimax, gunzburger2004uniformity}, and such criteria suggest that Optimal Latin hypercube sampling \cite{park1994optimal} and Sobol sequences \cite{niederreiter1988low, sobol1967distribution} offer a great balance between uniform and random sampling. In this work, we have chosen Sobol sequences (low discrepancy quasirandom sequences), to sample the input space. The main advantage of this technique is that the samples generated via this procedure are spread out over the input variables space non-uniformly, but cover the space evenly \cite{bessa2017framework}, thus allowing efficient exploration of the input space. An additional benefit is that as the Sobol sequence progresses, the input variables space is refined successively. This latter feature allows us to add simulations to the training in a systematic manner, till the desired accuracy is achieved in the ML model. Further details of the sampling procedure used are provided in Appendix \ref{Appendix:dgc}. 
% We populate the input space using the samples obtained from Sobol sequence. These samples are generated in sets which are space-filling by themselves, refine the input space successively, and sample in regions unexplored set and check the accuracy of the trained model. We continue adding simulations to the training from subsequent sets till the desired accuracy is achieved in the ML model. A decrease in the error as we cumulatively use these three sets to train our ML model is given in Appendix Fig. \ref{fig:error_sobolpca}. This scheme is helpful as it does not require you to know a \emph{priori} how many sample points in the input space would lead to an accurate ML model. Details regarding the input space are given in the Appendix \ref{Appendix:dgc}. 
% It is still unfeasible to simulate at all these grid points due to the high computational cost of these simulations; therefore, we simulate at only 204 data points. These 204 data points were selected using a Sobol sequence, which gives low discrepancy quasi-random sequences for sampling. Sobol sequences provide an advantage over the random generation of data points by generating data points more uniformly over the given space. Details regarding the grid in the input space are given in the Appendix \ref{Appendix:dgc}. 
\subsection{Dimensionality Reduction of the Electronic Fields and  Regression in the  Reduced  Dimension}
\label{subsec:dim_red_pca}
\underline{Dimensionality Reduction of the Electronic Fields:} We reduce the high dimensionality of the electronic fields using Principal Component Analysis (PCA) \cite{wold1987principal,jolliffe2002springer,lee2007nonlinear, jolliffe2016principal}. {PCA reduces the dimensionality of the data in an unsupervised manner.} PCA reduces the dimensionality of the data by projecting it onto a lower--dimensional space
such that the maximum statistical information within the data is retained. The basis vectors for this low--dimensional space are uncorrelated with each other and are called the principal components. Thus, PCA enables dimensionality reduction while minimizing the information loss. 
% ******************************************************The \emph{Karhunen-Lo$\grave{\text{e}}$ve} transform or Principal Component Analysis (PCA) \cite{wold1987principal,jolliffe2002springer,lee2007nonlinear, jolliffe2016principal} is a widely known  technique used for applications such as dimensionality reduction, data visualization and feature extraction. 
% PCA reduces the dimensionality of the data in an unsupervised manner by computing the linear projection of the data onto a new lower--dimensional space, such that the variance of the projected data is maximized. 

To elaborate further, given the data points , PCA allows one to obtain a lower dimensional approximation $\tilde{\mathbf{x}}_i\, \in \, \mathbb{R}^K$, such that, $K<d$, and:
\begin{equation}\label{eqn:PCA}
    \tilde{\mathbf{x}}_i = \sum_{j=1}^{K} c_{ij} \mathbf{v}_j + \bm{\mu}\,.
\end{equation}
Here, $\displaystyle \bm{\mu} = \frac{1}{n} \sum_{i=1}^{n}\mathbf{x}_i$ is the sample mean, the orthonormal vectors $\mathbf{v}_j \,$ are the principal components (PCs) and the scalars $c_{ij}\,$ are the coefficients of the principal components (CoPCs). Importantly, the PCs ($\mathbf{v}_j$) depend on the entire dataset rather than being associated with a particular data point; therefore, all the points in the original dataset can be defined in terms of distinct $c_{ij}\,$ values, but the same $\mathbf{v}_j$. The value of $K$ depends on the degree of variance of the data that needs to be captured. We perform PCA on the electronic field ($\rho$ and $b$) -- data generated by the DFT simulations. {Specifically, in Helical DFT, the discretized grid of the fundamental domain of electronic fields has $89$ points along the radial direction, $19$ points along the $\theta_1$ direction and $35$ points along the $\theta_2$ direction. Given this discretization, the total number of grid points in the fundamental domain is $59,\!185$. The electronic fields are therefore represented as a nearly $60,\!000$ dimensional vector by the DFT simulations. PCA enables us to represent these high-dimensional electronic fields in terms of a few CoPCS only ($7$ and $15$ for $\rho$ and $b$, respectively, as described in Section \ref{sec:results_pca_NN}).}

\underline{Regression for the  Electronic Fields in the Reduced Dimension:} We employ Neural Networks (NN) \cite{bishop2006pattern, wang2003artificial} to perform regression for the electronic fields in the reduced dimension. The choice of NN is motivated by our previous work on the ML based modeling of complex rippling deformation fields of low-dimensional nanostructures \cite{yadav2021interpretable}. The NN architecture consists of the input layer, multiple hidden layers, and the output layer. The neurons of the hidden layers contain a weighted linear transform of neurons in the previous layer acted upon by a nonlinear activation function. During the training phase of the model, the neural network learns {(in a supervised manner)} the map between input and output spaces by finding the weights of these linear transforms such that it can accurately predict output for a given input. We use NNs to predict the coefficients of the principal components (CoPCs) of the electronic fields for a given system and loading parameters. We deploy two different neural networks $\mathcal{N}_1$ and $\mathcal{N}_2$ to predict CoPCs for $\rho$ and $b$ respectively, which use the same input parameters. 
% \red{The neural networks $\mathcal{N}_1$ and $\mathcal{N}_2$ are trained in a supervised manner.} 
In the input layer, we have three neurons, for the \underline{input} parameters (i) $R_{\mathrm{avg}}$ , (ii) $\alpha$ and (iii) $\tau$. The neurons of the \underline{output} layer  correspond to the CoPCs ($c_{ij}$). Note that the number of CoPCs depends on the desired variance to be captured in the data. 

Inference via the trained ML model involves the following two steps. First, the CoPCs are predicted for a given input using the neural network. Second, the predicted CoPCs are used to obtain the higher dimensional electronic fields using the principal components following Eq. \ref{eqn:PCA}. These two steps inference procedure via the ML model are shown in Fig. \ref{fig:model_schematic}.
\subsection{Prediction of Nuclear Coordinates from Pseudocharge Fields}
\label{subsec:DBSCAN}
As mentioned earlier, calculation of the Kohn-Sham Hamiltonian arising from ML predicted fields requires knowledge of the nuclear coordinates so that the non-local part of the pseudopotential operator may be constructed (see Section \ref{subsec:Helical_DFT}). In this section we deal with the problem of obtaining these coordinates as a function of the tube geometry and loading parameters, i.e., $\{R_{\mathrm{avg}},\alpha,\tau\}$. 

One possible approach \citep{shi2021machine} is to directly train a neural network with these parameters as inputs and the desired nuclear coordinates as outputs. In our experience, however, this approach does not appear to work particularly well (see Appendix \ref{app:comparison_nn_dbscan}), and the amount of training data that was found to be adequate for predicting the electronic fields $\rho$ and $b$ accurately, was found to result in unacceptable levels of error while predicting the nuclear coordinates. This led us to devise a new strategy for determining the nuclear coordinates from the ML predicted nuclear pseudocharge field $b(\bfx)$, since this field is readily predicted with relatively high accuracy (Section \ref{subsec:electronic_fields}), and it already contains the nuclear coordinate information in principle. {Our proposed strategy to find nuclear coordinates is based on a clustering approach and identifies the nuclear coordinates from the nuclear pseudocharge field in an unsupervised manner.}

We make the observation that the nuclear pseudocharge field over the fundamental domain is a superposition of the individual atomic pseudocharges i.e., $\displaystyle b(\bfx) = \sum_{i=1}^{M} b_i(\bfx)$. Furthermore, each atomic pseudocharge field is spherically symmetric and atom centered (i.e. $b_i(\bfx) \equiv b_i(\lvert\bfx - \bfr_i\rvert)$), and under usual circumstances, also non-overlapping. This suggests that a clustering based approach that can identify agglomerations of positive charges arising from individual atoms might be fruitful, and the desired nuclear coordinates can then be determined as cluster centers. Clustering algorithms are widely employed  to divide datasets into smaller subgroups in an unsupervised manner, such that the data points in each subgroup share some common attributes \cite{bishop2006pattern}. 
%\red{************** Combine Paragraphs}
 %\blue{Here, we utilize density-based spatial clustering of applications with noise (DBSCAN) procedure on the nuclear pseudocharge field $b(\bfx)$ . DBSCAN clusters the datapoints such that points that are closely packed together are grouped \cite{ester1996density, schubert2017dbscan}.}
One of the most widely used and successful clustering algorithms is DBSCAN (Density-Based Spatial Clustering of Applications with Noise) \cite{ester1996density, schubert2017dbscan}. This technique creates clusters for volumes with a high density of points, and treats the points which lie in very low-density volumes as outliers. DBSCAN offers advantages over other clustering algorithms like $k-$nearest neighbors, since it does not require prior knowledge of the number of clusters present in the data, it can find out any arbitrarily shaped clusters and it is robust against errors induced by the outliers. In the present case, this means that when applied to the nuclear pseudocharge data, DBSCAN should be able to form clusters around every nucleus in the fundamental domain, without the total number of nuclei being specified apriori. However, we found that a direct application of DBSCAN to the pseudocharge field fails to determine the nuclear coordinates accurately. In the following, we identify two reasons for this failure and develop procedures to overcome them. 

\begin{figure}[htbp]
    \centering
    \includegraphics[width=0.3\linewidth]{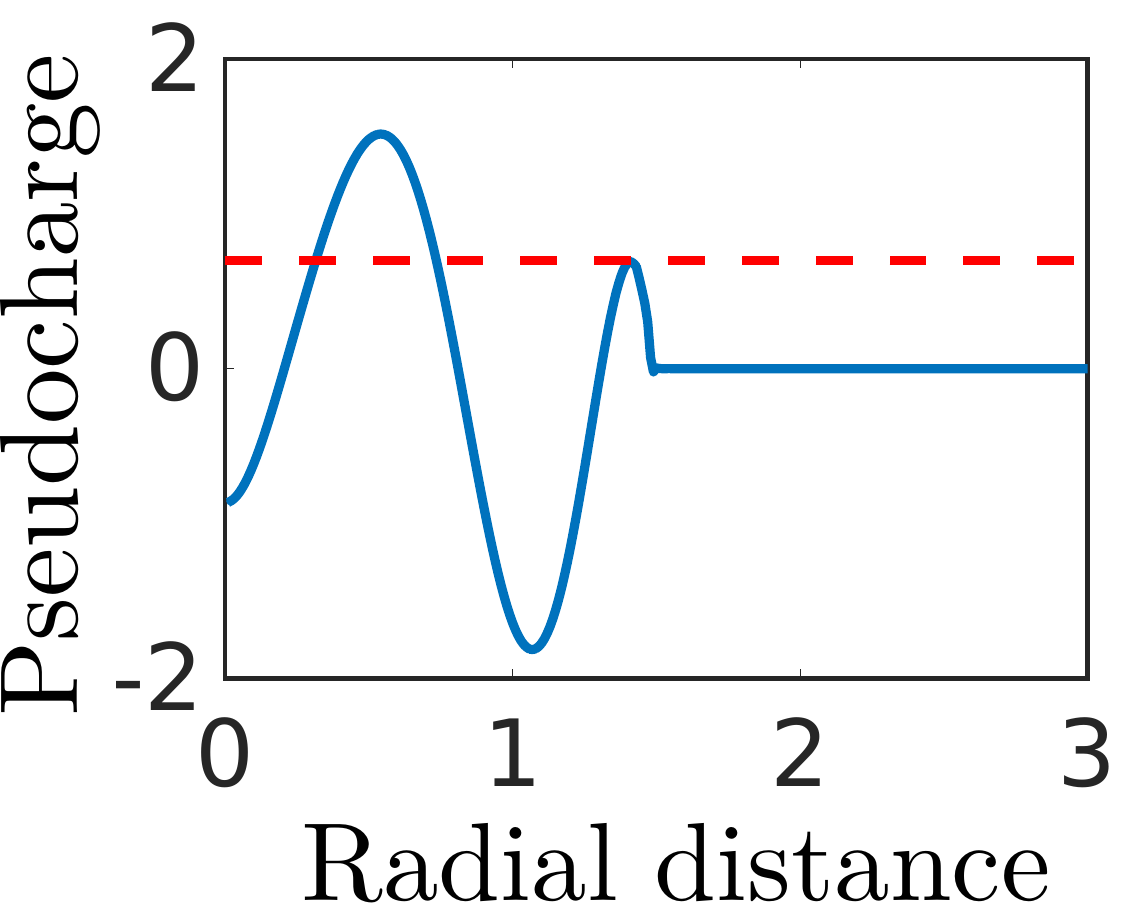}
    % \caption{Cluster formation from nuclear pseudocharge field to determine nuclei position. Red clusters show the positive charge around nucleus and the black dots are nuclei.}
    \caption{Atomic pseudocharge as a function of distance (in Bohr) from the atom for the Troullier-Martins pseudopotential for Carbon used in this work. The dashed red line indicates the truncation level employed before the DBSCAN procedure is used. }
    \label{fig:b_vs_dist}
\end{figure}

First, the fundamental domain is effectively periodic in the $\theta_1$ and $\theta_2$ directions. However, clustering algorithms are not typically aware of domain boundary conditions, as a result of which, pseudocharges associated with atoms close to the domain edges may result in the identification of clusters for which the cluster centers are not at the nuclear coordinates. This issue is readily addressed by expanding the fundamental domain into a supercell, applying the clustering procedure to the periodically replicated pseudocharge field in the supercell, and finally, retaining the cluster centers found to lie within the fundamental domain. Second, some atomic pseudocharges (such as the one associated with the Troullier Martins pseudopotential for Carbon used in this work), while being radially symmetric, may exhibit multiple peaks, when plotted as a function of atom center distance (see Figure \ref{fig:b_vs_dist}). This can cause the clustering algorithm to identify multiple clusters near a single nucleus and the centers of these clusters will not coincide with the nuclei. To overcome this challenge we propose a map ($\mathcal{T}$) that truncates the pseudocharge field $b$ to retain only the data around the first peak (see Figure \ref{fig:b_vs_dist}):
\begin{equation}
    \mathcal{T} : b (r,\theta_1,\theta_2) \rightarrow  \bar{b} (r,\theta_1,\theta_2),\;\;  \bar{b} (r,\theta_1,\theta_2) = 
    \begin{cases}
        b (r,\theta_1,\theta_2),& \text{if } b (r,\theta_1,\theta_2) > c_t\\
        0,              &  \text{if } b (r,\theta_1,\theta_2) \leq c_t
    \end{cases}
\end{equation}
The only quantitative information needed for implementing this map is the height of the second peak $c_t$, which is readily available for the pseudopotentials used to produce the training data. The DBSCAN procedure, when applied on the truncated field $\bar{b}$ can readily identify the nuclear pseudocharge density cluster around each nucleus. Nuclear coordinates are subsequently computed as centers of these clusters. 

Together, the above set of strategies leads to a robust and efficient method for obtaining the nuclear coordinates as a function of the ML model inputs. The entire procedure outlined above executes within a few seconds of wall time on a desktop and is able to %\blue{pick out} 
{determine} the nuclear coordinates to acceptable levels of accuracy in every case (see Table \ref{tab:Energy_Errors}). Comparison of the accuracy of our clustering based approach, with that of nuclear coordinate predictions using a standard neural network are presented in Appendix \ref{app:comparison_nn_dbscan}.
\begin{figure}[htbp]
    \centering
    \includegraphics[width=0.95\linewidth]{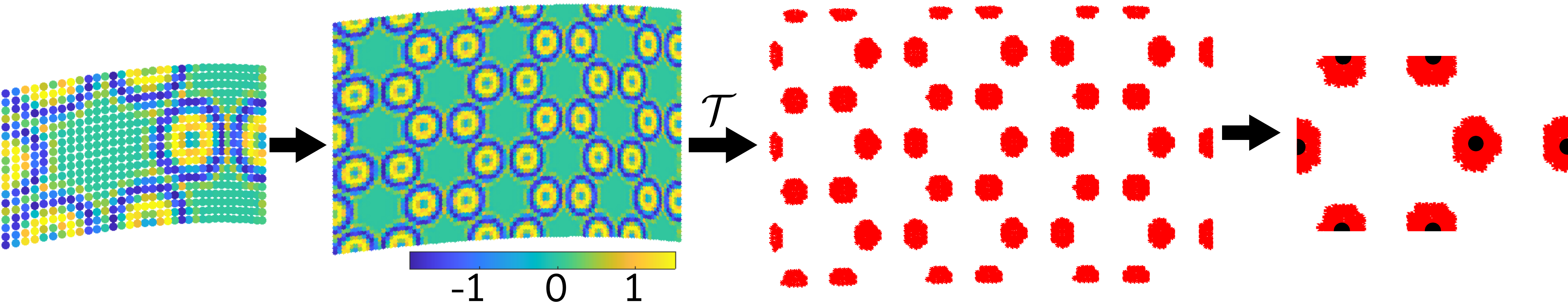}
    % \caption{Cluster formation from nuclear pseudocharge field to determine nuclei position. Red clusters show the positive charge around nucleus and the black dots are nuclei.}
    \caption{Cluster formation from nuclear pseudocharge field to determine nuclei position. {A slice of the pseudocharge field at the average radial coordinate of the atoms in the fundamental domain is shown}. Red clusters show the positive charge around the nucleus and the black dots are nuclei. The pseudocharge field on the fundamental domain is expanded to a supercell to avoid domain edge effects, a truncation is implemented to discard secondary peaks in the atomic pseudocharges, the DBSCAN procedure is then applied on the supercell and finally, the nuclear coordinates within the fundamental domain are identified.}
    \label{fig:clustering}
\end{figure} 
\section{Post-processing of ML Predicted Electronic Fields}
\label{sec:postprocessing}
In this section, we describe the postprocessing steps used for computing quantities of interest from the machine learning model predicted fields and atomic coordinates. The machine learning model produces electronic fields $\rho^{\text{ML}}(\bfx)$ and $b^{\text{ML}}(\bfx)$ that includes self-consistency and atomic relaxation effects. Within the ML model, however, we do not explicitly enforce any constraints regarding the net charges associated with these fields. Although in practice these constraints seem to be automatically obeyed by the model --- at least approximately (see Table \ref{tab:ne} within the section on Results), we find it useful to scale the ML predicted fields for postprocessing purposes \cite{ALRED20183}, as shown below: 
\begin{align}
\nonumber
\rho^{\text{Scaled}}(\bfx) &= \rho^{\text{ML}}(\bfx) \times \frac{N_{\text{e}}} {\int_{\calD} \rho^{\text{ML}}(\bfx)\,d\bfx}\,,\\
b^{\text{Scaled}}(\bfx) &= b^{\text{ML}}(\bfx) \times \frac{-N_{\text{e}}} {\int_{\calD} b^{\text{ML}}(\bfx)\,d\bfx}\,.
\label{Eqn:rescale}
\end{align}
Using these scaled fields, we compute the net electrostatic potential $\Phi$ via iterative solution of eq.~\ref{Eqn:poisson} using preconditioned GMRES \citep{saad1986gmres} iterations. The exchange correlation potential $V_{\text{xc}}$ is directly computed from the electron density. Next, we use a clustering based unsupervised learning technique (see Section \ref{subsec:DBSCAN}) to pick out the nuclear coordinates from the nuclear pseudocharge field and use it to set up the non-local pseudopotential operator $\mathcal{V}_{\text{nl}}$. Thereafter, we diagonalize the Kohn-Sham Hamiltonian (eq.~\ref{Eqn:ks_eqn}) resulting from these machine learning predicted quantities, to obtain the Kohn-Sham eigenstates. We use a combination of Generalized Preconditioned Locally Harmonic Residual (GPLHR) \citep{vecharynski2015generalized} and Arnoldi Iterations \citep{Saad_large_eigenvalue_book} to carry out the diagonalization, and initialize the calculations using random wavefunction vectors. The Fermi level of the system is subsequently determined from the Kohn-Sham eigenvalues by enforcing the constraint of having a fixed number of electrons within the fundamental domain. {From all this information, band structure dependent quantities, such as the value of the band gap and its location in $(\eta, \nu)$ space, can be calculated.}

Using the aforementioned post-processed quantities, the ground state free energy per unit fundamental domain may be calculated as \citep{yu2022density}:
\begin{align}
\mathcal{F} = E_{\text{kin}} + E_{\text{xc}} + E_{\text{nl}} + E_{\text{el}} - T_{\text{e}}S\,,
\label{Eqn:total_nrg}
\end{align}
with the terms on the right hand side denoting the electronic kinetic energy, the exchange correlation energy, the non-local pseudopotential energy, the electrostatic energy and the electronic entropy contribution at temperature $T_\text{e}$, respectively. Alternately, a more accurate estimate for the ground state free energy per unit fundamental domain may be obtained using the Harris-Foulkes functional \citep{harris1985simplified, foulkes1989tight}:
\begin{align}
\label{eq:HF}
{\calF}^{\text{HF}} = E_{\text{band}} + E_{\text{xc}} - \widetilde{E}_{\text{xc}} + \widetilde{E}_{\text{el}} + E_{\text{sc}} - T_{\text{e}}\,S\,.
\end{align}
In the above, the first term on the right hand side is the electronic band energy:
\begin{align}
 E_{\text{band}} = 2\int_{-\half}^{\half}\frac{1}{\mathfrak{N}}\sum_{\nu = 0}^{\mathfrak{N}-1}\sum_{j=1}^{\infty}\lambda_j(\eta, \nu)\,g_j(\eta, \nu)\,d\eta\,,
\end{align}
in which $\displaystyle g_j(\eta,\nu)$ denotes the electronic occupations. The term  $E_{\text{xc}}$ denotes the exchange correlation energy, while: 
\begin{align}
\widetilde{E}_{\text{xc}} = \int_{ {\calD}}V_{\text{xc}}(\rho(\bfx))\rho(\bfx)\,d\bfx\,.
\end{align}
The term $\widetilde{E}_{\text{el}}$ is related to electrostatic interactions and has the form:
\begin{align}
\widetilde{E}_{\text{el}} = \half\int_{{\calD}}\big(b(\bfx) - \rho(\bfx) \big)\Phi(\bfx)\,d\bfx\,.
\end{align}
Finally, $E_{\text{sc}}$ accounts for nuclear pseudocharge self-interactions and overlap corrections \citep{banerjee2016cyclic}, while the last term is related to the electronic entropy contribution. Notably, in the above breakdown for the Harris-Foulkes energy, $E_{\text{xc}}$ and $\widetilde{E}_{\text{xc}}$ depend solely on the electron density field, $E_{\text{sc}}$ depends on the nuclear coordinates and the nuclear pseudocharge field, the electrostatic term $\widetilde{E}_{\text{el}}$ depends on both the electron density and nuclear pseudocharge fields, while $E_{\text{band}}$ depends on the Kohn-Sham operator eigenvalues (i.e., its dependence on $\rho$ and $b$ is implicit). Therefore, monitoring these terms in addition to  ${\calF}^{\text{HF}}$ allows us to estimate the accuracy of the machine learning based predictions of $\rho$, $b$ and other post-processed quantities (such as the eigenstates), in the energetic sense (see Section \ref{sec:Results} for more details).

%~~~~~~~~~~~~~~~~~~~~~~~~~~~~~~~~~~~~~~~~~~~~~~~~~~~~~~~~~~~~~~~~~~~~~~~~~~~~~~~~~~~~~~~~~~~~~~~~~~~~~~~~~~~~~
\section{Results}\label{sec:Results} 
We now present the predictions of the machine learning (ML) model for armchair carbon nanotubes under torsional and axial loading. These are compared against Helical DFT simulations to quantify the ML model's accuracy and efficacy. Notably, the inference process from the trained ML model is orders of magnitude faster compared to the cost of the ab initio simulations using Helical DFT. While the ML model requires $0.003$ seconds and $0.009$ seconds to predict the $\rho$ and $b$ fields respectively (average times on a desktop with a $2.2$ GHz Intel Xeon Gold processor), a typical ab initio structural relaxation calculation using Helical DFT can stretch into hundreds of CPU hours. Post-processing of the ML predicted electronic fields (to calculate band structures, energies, etc.) can be typically performed in about $30$ to $40$ minutes of wall time. Training of the neural networks for $\rho$ and $b$ requires about $12$ and $15$ minutes, respectively, measured using the same hardware setup. %\blue{As expected, the ML model is significantly faster than the ab initio simulations, even with the use of a specialized and efficient first principles simulation technique.}  

\subsection{Principal Component Analysis and Neural Networks} 
\label{sec:results_pca_NN}
\begin{figure}[htbp]
    \centering
    \includegraphics[width=0.49\linewidth]{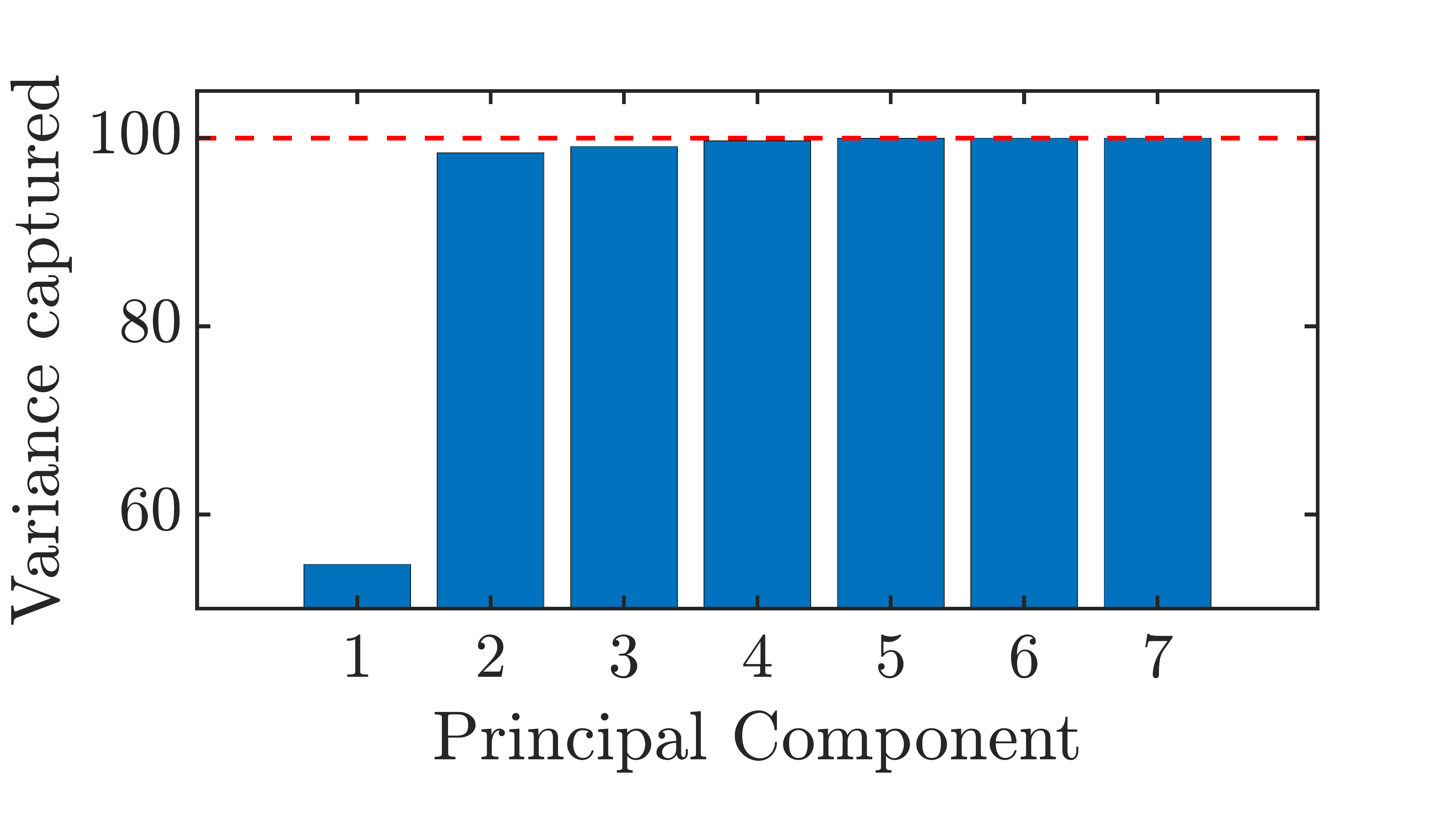}
    \includegraphics[width=0.49\linewidth]{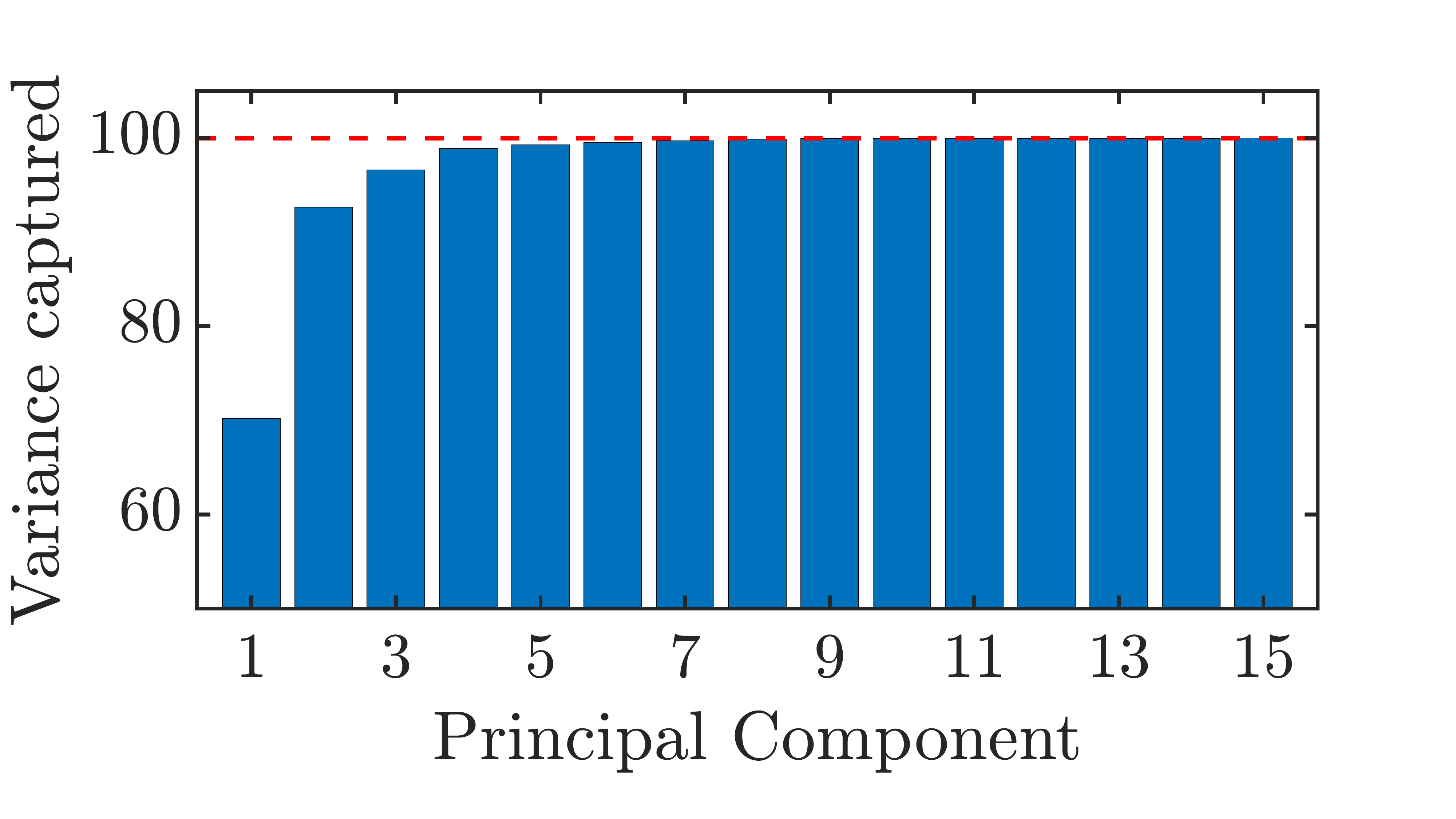}
    \caption{Cumulative percentage of variance vs Principal components for $\rho$ (\emph{left}) and $b$ (\emph{right}). The red dashed line shows $99.99$\% variance. }
    \label{fig:pca_var}
\end{figure}

\underline{Principal Component Analysis Results:} As the first step in our two-step ML model, we utilize PCA to obtain reduced dimensional representations for the outputs of the map $\mathcal{H}$. To reconstruct the original electronic fields with minimum reconstruction error, we capture $99.99$\% variance of the data. As shown later (Section \ref{subsec:post_proc_results}), this is generally sufficient for obtaining electronic ground state energies to chemical accuracy and also adequate for reproducing band structures correctly. For capturing this level of variance in the data, we required only $7$ PCs in case of $\rho$ and $15$ PCs in case of $b$ (Figure \ref{fig:pca_var}).

\underline{Neural Network:} As the second step in our two-step ML model, two Neural Networks $\mathcal{N}_1$ and $\mathcal{N}_2$ are trained to predict CoPCs corresponding to $\rho$ and $b$, respectively. Since $7$ PCs in case of $\rho$ and $15$ PCs in case of $b$ are required to capture 99.99\% variance of the data, the number of neurons in output layers is 7 for $\mathcal{N}_1$ and 15 for $\mathcal{N}_2$. Following our architecture optimization strategy (elaborated in Appendix \ref{Appendix:DNN_training}) we choose $6$ hidden layers of $150$ neurons each for $\mathcal{N}_1$ and $2$ hidden layers of $150$ neurons each for $\mathcal{N}_2$.  We use Rectified Linear Unit (ReLU) as an activation function for both networks. Mean Squared Error(MSE) is utilized as a loss function along with the elastic net regularization \cite{zou2005regularization}, and the Adam optimizer \cite{kingma2014adam} with a learning rate of $0.001$ was employed. Before the training phase, each input parameter column was scaled to zero mean and unit variance, thus standardizing the input features. $75\%$ of the total data points were utilized for training ($123$ data points), $10\%$ were utilized for validation ($16$ data points), and the remaining $15\%$ were utilized for testing (25 data points).  Further details of the neural network, including a discussion of the hyperparameters, and learning curves are provided in Appendix \ref{Appendix:DNN_training}.  

\subsection{Prediction of electronic fields by the ML model} 
\label{subsec:electronic_fields}
\begin{figure}[htbp]
    \centering
    \includegraphics[width=0.99\linewidth]{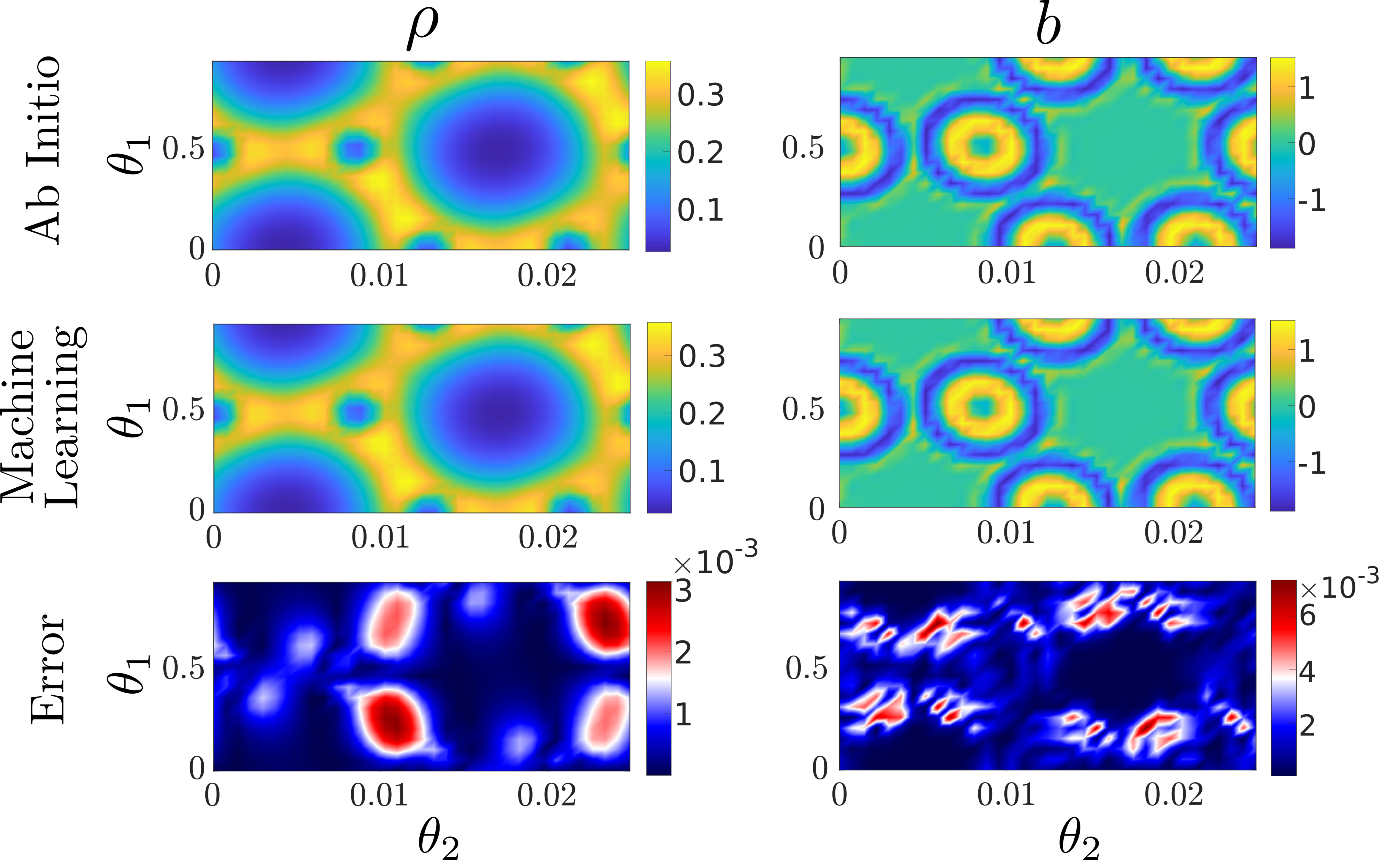}
    \caption{Comparison between ML predicted and DFT simulation obtained electronic fields for a test data point with all unknown input parameters ($R_{\mathrm{avg}} = 49.51 \, \text{Bohr}$, $\alpha = 0.00125$, $\tau = 4.5552 \, \text{Bohr}$). {A slice of the electronic fields at the average radial coordinate of the atoms in the fundamental domain is shown}. The error is computed as $\frac{\vert \rho^{\text{DFT}} - \rho^{\text{ML}} \vert}{\vert \mathrm{max}(\rho^{\text{DFT}}) - \mathrm{min}(\rho^{\text{DFT}})\vert}$, similarly for $b$. Here, max($\cdot$) and min($\cdot$) denote maximum and minimum over the fundamental domain.}
    \label{fig:fields}
\end{figure}

We now discuss the overall performance of the machine learning model for the prediction of the electronic fields. The Pearson correlation coefficient ($R$) between the predicted and actual electronic fields at each point of the discretized domain for the test data was found to be $0.9949$ and $0.9983$ for $\rho$ and $b$, respectively. %(i.e., $\rho (r_i,{(\theta_1)}_j,{(\theta_2)}_k)$ and $b(r_i,{(\theta_1)}_j,{(\theta_2)}_k)$, for $i = 1,2,\ldots,\mathsf{N}_r ;\, j = 1,2,\ldots, \mathsf{N}_{\theta_1};\, k = 1,2,\ldots, \mathsf{N}_{\theta_2}$)%, 
\begin{figure}[htbp]
    \centering
    \includegraphics[width=0.7\linewidth]{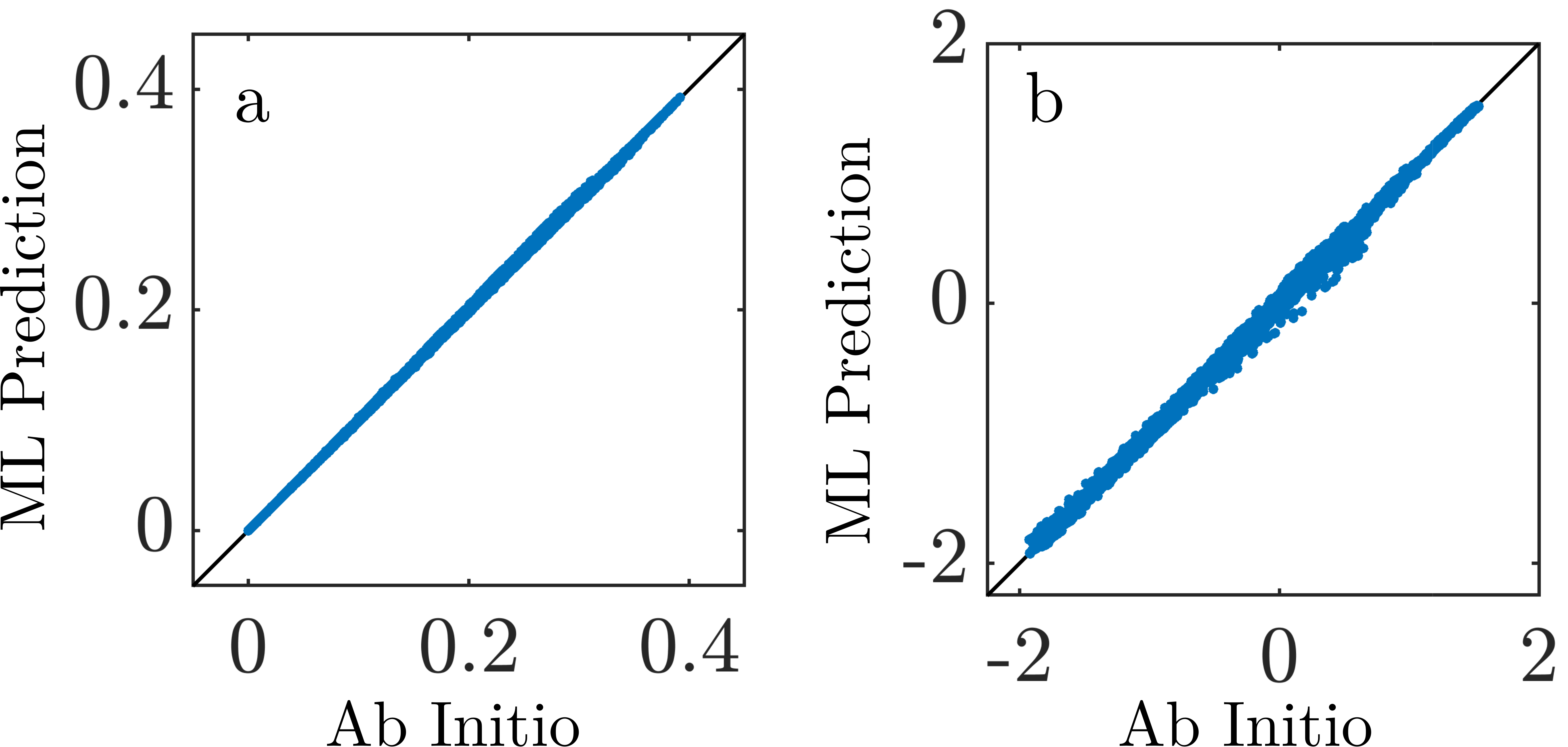} %use parity.pdf for 4 parity plots
    \caption{Parity plots for (a)\,test data of $\rho$ ($R=0.9949$), (b) test data of $b$ ($R=0.9983$).}
    \label{fig:parity}
\end{figure} 
In addition to the test data points described above, we have chosen three additional test data points where the input parameters were partially unseen during the training (i.~$R_{\mathrm{avg}} = 49.51 \, \text{Bohr}$, $\alpha = 0.002$, $\tau = 4.5052 \, \text{Bohr}$; ii.~$R_{\mathrm{avg}} = 35.55 \, \text{Bohr}$, $\alpha = 0.00125$, $\tau = 4.6052 \, \text{Bohr}$; iii.~$R_{\mathrm{avg}} = 53.32 \, \text{Bohr}$, $\alpha = 0.0015$, $\tau = 4.5552 \, \text{Bohr}$). For each of these three test cases, there is one input variable whose value was not used in the training data (e.g. data point with $\tau$ and $\alpha$ values present in the training data but the value of $R_{\mathrm{avg}}$ not present in the training data). %That is, either the system or one of the loading parameters is not known. 
Finally, we have randomly selected two additional test data points where none of the three input variables were seen by the ML model during training (i.~$R_{\mathrm{avg}} = 49.51 \, \text{Bohr}$, $\alpha = 0.00125$, $\tau = 4.5552 \, \text{Bohr}$; ii.~$R_{\mathrm{avg}} = 30.46 \, \text{Bohr}$, $\alpha = 0.00075$, $\tau = 4.6552 \, \text{Bohr}$). %That is, neither the system, nor the loading are known. 
These additional test data points with partial or wholly unseen input parameters help assess the ML  model's capability to generalize beyond training data. Machine Learning predicted and actual (DFT) electronic fields for one of the test data points with all unknown input parameters are compared in Fig. \ref{fig:fields}. 

We quantify the error in the predicted electronic fields through the normalized root mean square error (NRMSE) \cite{shi2021machine}: %computed as root mean squared error over all grid points ($89 \times 35 \times 19$) normalized by the absolute of range of the actual electronic field values. 
\begin{equation}
    \text{NRMSE} = \frac{\sqrt{\frac{1}{d}\sum_{i=1}^d \left(\rho_i^{\text{DFT}} - \rho_i^{\text{ML}}\right)^2}}{\vert \text{max}(\rho^{\text{DFT}}) - \text{min}(\rho^{\text{DFT}}) \vert}
\end{equation}
Here max($\cdot$) and min($\cdot$) denote maximum and minimum over the fundamental domain and $d$ is the dimension of the data ($\sim 60,000$). NRMSE for $b$ is calculated similarly. The NRMSE for various categories of test data points, including cases with partial or wholly unseen inputs, are presented in Table \ref{tab:ne}. 
The low NRMSE values on the test data are indicative of the general accuracy of the ML model. In particular, low NRMSE values for the input conditions beyond the training data establish the generalization capacity of the model. 
% We approximate the electronic fields through CoPCs. These CoPCs, along with PCs, can reconstruct the high-dimensional electronic field. To limit the number of CoPCS, we capture 99.99\% variation in data, which leads to an error in reconstruction of high--dimensional electronic field. Note that PCs are obtained using only training data (i.e., 122 data points). NRMSE (averaged over 122 training data points) in reconstruction of $\rho$ and $b$ were found to be $ 4.5231 \times 10^{-5} $ and $ 1.2682 \times 10^{-4}$. We can therefore say that CoPCs accurately represent the electronic fields. \\

In addition to evaluating the NRMSE values, we monitored the integrals of $\rho^{\text{ML}}$ and $b^{\text{ML}}$ over the fundamental domain. For a neutral system with $N_{\text{e}}$ electrons within the computational unit cell, the electron density and the nuclear pseduocharge fields obey the normalization conditions $\displaystyle \int_{\calD} \rho ( \mathbf{x} )\,\mathrm{d}\mathbf{x} = N_{\text{e}}$ and $\displaystyle \int_{\calD} b(\mathbf{x})\,d\mathbf{x} = -N_{\text{e}}$ respectively. Since these constraints were not built into the ML model, they allow additional quality checks on the ML predicted fields to be performed. As shown in Table \ref{tab:ne}, the errors associated with deviations from these constraints are quite low ($0.000625$ particles or lower, per electron), indicating high quality predictions of the electronic fields by the ML model.
\begin{table*}[htbp]
\resizebox{\textwidth}{!}{%
\begin{tabular}{|c|c|c|c|c|}
\hline
% Case & NRMSE ($\rho$) & $\vert \int \rho^{DFT} ( \mathbf{r} ) \mathrm{d}^3\mathbf{r} - \int \rho^{ML} ( \mathbf{r} ) \mathrm{d}^3\mathbf{r}\vert$ & NRMSE ($b$) & $\vert\int b^{DFT} ( \mathbf{r} ) \mathrm{d}^3\mathbf{r} - \int b^{ML} ( \mathbf{r} ) \mathrm{d}^3\mathbf{r}\vert$ \\ \hline
Case & NRMSE ($\rho$) & $\big\vert N_{\text{e}}  - \int_{\calD} \rho^{\text{ML}} ( \mathbf{x} ) \mathrm{d}\mathbf{x}\big\vert$ & NRMSE ($b$) & $\big\vert (-N_{\text{e}})  - \int_{\calD} b^{\text{ML}} ( \mathbf{x} ) \mathrm{d}\mathbf{x}\big\vert$ \\ \hline
Average for test data set & $2.8 \times 10^{-4}$ & $3.2 \times 10^{-3}$ & $5.6 \times 10^{-4}$ & $5.4 \times 10^{-3}$ \\ \hline
\begin{tabular}[c]{@{}c@{}}Random test data point\\ $R_{\mathrm{avg}} = 40.64\; \text{}, \alpha = 0.0015, \tau = 4.7052\;\text{}$\end{tabular} &  $2.8 \times 10^{-4}$ & $9.2 \times 10^{-3}$  & $2.8 \times 10^{-4}$  & $8.6 \times 10^{-3}$ \\ \hline
\begin{tabular}[c]{@{}c@{}}Test data point with unknown $R_{\mathrm{avg}}$\\ $R_{\mathrm{avg}} = 49.51\; \text{}, \alpha = 0.002, \tau = 4.5052\;\text{}$\end{tabular} & $1.3 \times 10^{-4}$ & $5.2 \times 10^{-4}$ & $2.1 \times 10^{-4}$ &  $2.5 \times 10^{-3}$ \\ \hline
\begin{tabular}[c]{@{}c@{}}Test data point with unknown $\alpha$\\ $R_{\mathrm{avg}} = 35.55\; \text{}, \alpha = 0.00125, \tau = 4.6052\;\text{}$\end{tabular} & $2.5 \times 10^{-4}$ & $1.2 \times 10^{-3}$ & $1.9 \times 10^{-4}$ &  $5.6 \times 10^{-3}$ \\ \hline
\begin{tabular}[c]{@{}c@{}}Test data point with unknown $\tau$\\ $R_{\mathrm{avg}} = 53.32\; \text{}, \alpha = 0.0015, \tau = 4.5552\;\text{}$\end{tabular} & $1.6 \times 10^{-4}$ & $4.2 \times 10^{-3}$ & $2.5 \times 10^{-4}$ & $8.6 \times 10^{-3}$ \\ \hline
\begin{tabular}[c]{@{}c@{}}Test data point with unknown $R_{\mathrm{avg}}$, $\alpha$, $\tau$\\ $R_{\mathrm{avg}} = 49.51\; \text{}, \alpha = 0.00125, \tau = 4.5552\;\text{}$\end{tabular} & $2.1 \times 10^{-4}$ & $4.3 \times 10^{-3}$ & $4.8 \times 10^{-4}$ & $3.1 \times 10^{-3}$\\ \hline
\begin{tabular}[c]{@{}c@{}}Test data point with unknown $R_{\mathrm{avg}}$, $\alpha$, $\tau$\\ $R_{\mathrm{avg}} = 30.46\; \text{}, \alpha = 0.00075, \tau = 4.6552\;\text{}$\end{tabular} & $2.3 \times 10^{-4}$ & $3.0 \times 10^{-3}$ & $2.8 \times 10^{-4}$ & $2.8 \times 10^{-3}$\\ \hline
\end{tabular}
}
\caption{Table showing NRMSE for ML predicted $\rho$ and $b$ for various test cases. Also shown are errors in the integrals of electronic fields over the fundamental domain. $R_{\mathrm{avg}}$ and $\tau$ values are in Bohr.}
\label{tab:ne}
\end{table*}
% Average NRMSE for test data set was found out to be  $2.814 \times 10^{-4}$ and $5.570 \times 10^{-4}$ for $\rho$ and $b$, respectively. 
% NRMSE for data point with all unseen input parameters is $2.197 \times 10^{-4}$ and  $4.847 \times 10^{-4}$ for $\rho$ and $b$ respectively. We also assesed NRMSE values for another data point ($R_{\mathrm{avg}} = 30.46 \text{Bohr}$, $\alpha = 0.00075$, $\tau = 4.6552 \text{Bohr}$) in the input space whose input values are not seen during the training, and we observed low NRMSE on those data points as well.
\subsection{Prediction of nuclear coordinates, energies and band structure}
\label{subsec:post_proc_results}
Finally, we post-process the ML predicted electronic fields for various test data points to obtain nuclear coordinates, electronic properties and energy components of interest. We compute the errors in these quantities for a random test data point, as well as the aforementioned five test cases for which the inputs were partially or wholly unseen by the ML model during training (Table \ref{tab:Energy_Errors}). In general, the errors in the total ground state energy, as computed through the Harris-Foulkes functional (eq.~\ref{eq:HF}) are found to be appreciably smaller than the chemical accuracy threshold ($1.6 \times 10^{-3}$ Ha/atom), except for one of the cases which had an unseen value of $\alpha$.  Considering the various components of the Harris-Foulkes energy, we see that the highest accuracies in the ML predictions are associated with the exchange correlation term ${E}_{\text{xc}}$, possibly due to the sole dependence of this quantity on the electron density, which itself is predicted rather accurately. The energy component $\widetilde{E}_{\text{xc}}$ (eq.~\ref{eq:HF}) also has a very similar behavior and is not shown in Table \ref{tab:Energy_Errors}. The nuclear self-energy and correction terms which depend only on the nuclear pseudocharge field are also predicted with high accuracy. The electrostatic term which depends on both the nuclear pseudocharge field and the electron density, and the electronic band energy, which depends on the Kohn-Sham eigenvalues are seen to be associated with somewhat lower accuracy predictions, particularly for the test data points which had values of $\alpha$ and/or $\tau$ unseen by the ML model. However, even in these cases, the errors are less than $3.0 \times 10^{-3}$ Ha/atom, and error cancellation leads to overall accurate ground state energy predictions. The ability to predict ground-state energies of deformed quasi-one-dimensional structures (while having atomic relaxation effects already included) with first principles accuracy, at a small computational cost is one of the great advantages of the proposed ML model, thus leading to its potential use in the multiscale modeling of low-dimensional systems \citep{hakobyan2012objective}.
      
\begin{figure*}[htbp]
\centering
\subfloat[Symmetry adapted band diagram in $\eta$, at $\nu = 2$.]{\scalebox{0.55}
{\includegraphics[width=\linewidth]{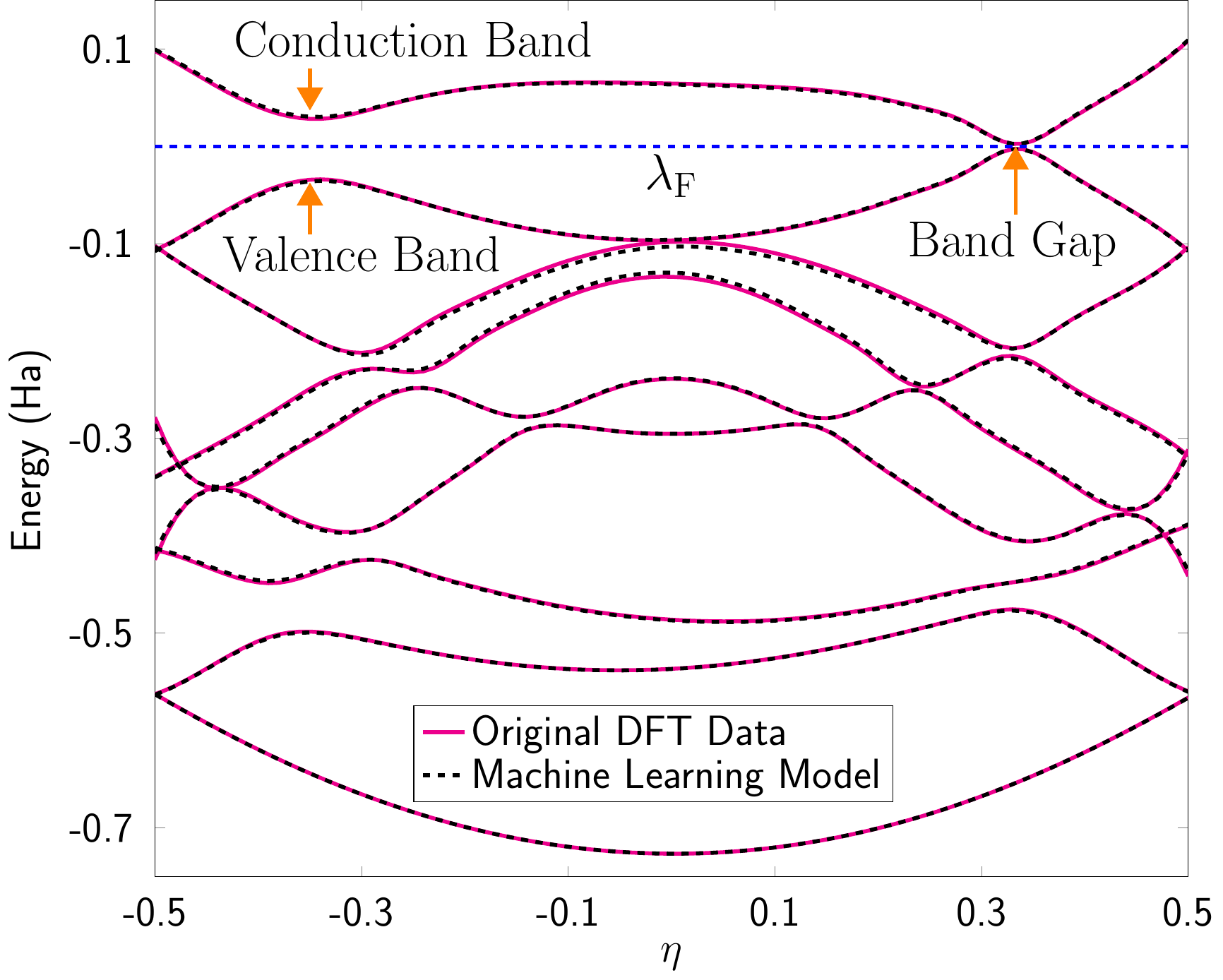}}}
\\
\subfloat[Symmetry adapted band diagram in $\nu$, at $\eta = 0$.]{\scalebox{0.55}
{\includegraphics[width=\linewidth]{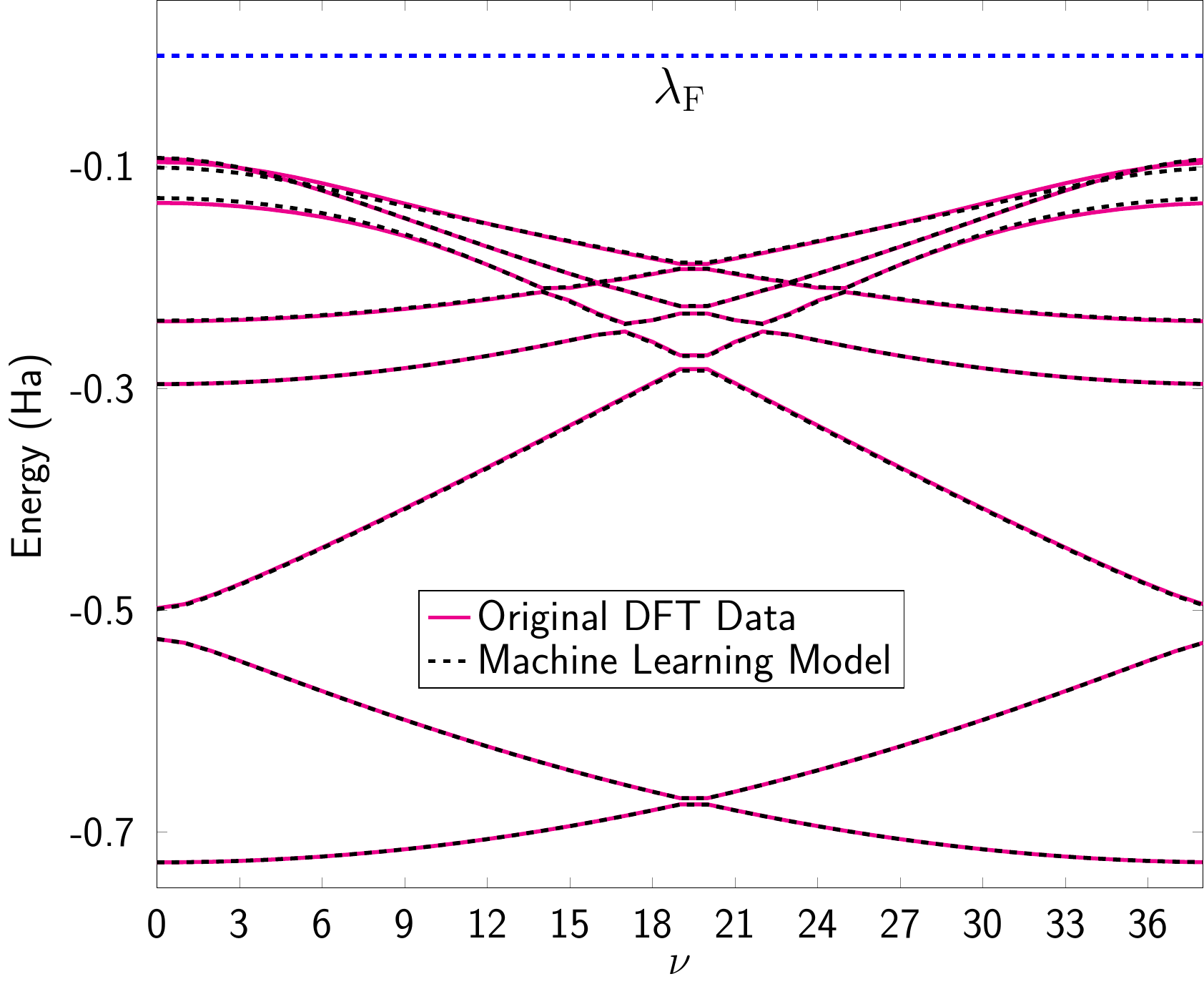}}}
\caption{Comparison of symmetry adapted band diagrams produced by the original DFT method and the machine learning model (with post processing) for the unknown test data point with $R_{\mathrm{avg}} = 49.51\;\text{Bohr}, \alpha = 0.00125$ and $\tau = 4.5552\;\text{Bohr}$.  The agreement appears excellent and the post--processed ML model is also able to precisely predict the location of the band--gap (at $\eta = \frac{1}{3}, \nu = 2$) as well as its value ($0.128$ eV from Helical DFT) to about $6\%$ accuracy in this case. Note that $\lambda_{\text{F}}$ denotes the system's Fermi level.}
\label{fig:electronic_states_plots}
\end{figure*}

The unsupervised learning procedure used for picking out nuclear coordinates is also found to be quite accurate, with typical errors (measured as the maximum error in the Cartesian coordinate components of all atoms in the fundamental domain) of the order of $0.02$ to $0.03$ Bohrs. The accuracy in the prediction of these coordinates is also reflected in the overall accuracy of the ML predicted Kohn-Sham Hamiltonian, which in turn, affects the quality of electronic band diagrams and other eigenstate-dependent quantities computed from the Hamiltonian. We found strikingly good agreement between ML predicted and Helical DFT band diagrams for the test data points considered here, with a typical case (associated with  wholly unseen inputs) demonstrated in Fig \ref{fig:electronic_states_plots}. Undeformed armchair carbon nanotubes are metallic \citep{ghosh2019symmetry, ding2002analytical} but develop an oscillatory band gap as a function of imposed twist \citep{ding2002analytical, yu2022density}. The band gap (computed here as the difference between the smallest eigenvalue above the Fermi level and the largest eigenvalue below the Fermi level as the symmetry indices $(\eta,\nu)$ are varied) is particularly error prone since it is the difference of two quantities. {Additionally, as the tube is deformed, the location and nature (i.e., direct vs. indirect) of the band gap is expected to change \citep{yu2022density}.} However, the ML predicted location of the band gap was correct for every test case and its value was correct to about $0.02$ eV or better, in almost every test case. {In this regard, our approach to predicting band structure dependent quantities by means of a post-processing step applied to the ML predicted electronic fields appears to be especially effective. In contrast, as shown in Appendix \ref{app:bandloc_nn_ourapproach}, direct prediction of such quantities using a neural network which takes in the input parameters $(R_{\mathrm{avg}},\alpha,\tau)$, can be error prone --- particularly, for the relatively small amount of data that was required for accurate prediction of the electronic fields themselves.} Overall, the ability of our approach to accurately and efficiently predict the electronic structure of low-dimensional materials as a function of imposed deformation, opens up the use of such techniques for strain-engineering applications \citep{dai2019strain, si2016strain, jiang2015strain}. 
\begin{table*}[htbp]
\resizebox{\textwidth}{!}{%
\begin{tabular}{ |c|c|c|c|c|c|c|c| } 
 \hline
 \multirow{4}{*}{Case} & {Ground state} & Exch.~Corr.  & Electrostatic & Nuclear self energy \& & Band  & Band & Atomic \\
 & energy & energy & term & correction term & Energy & gap & coordinates \\
 & ${\calF}^{\text{HF}}$ & $E_{\text{xc}}$ & $\widetilde{E}_{\text{el}}$  & $E_{\text{sc}}$ & $E_{\text{band}}$ &  & $\bfr_{i}$ \\
 & (Ha/atom) &  (Ha/atom) & (Ha/atom) & (Ha/atom) & (Ha/atom) & (eV) & (Bohr) \\\hline 
 Random test data point & \multirow{2}{*}{$9.0\times10^{-4}$} &\multirow{2}{*}{$4.1\times10^{-5}$} & \multirow{2}{*}{$7.1\times10^{-5}$} & \multirow{2}{*}{$1.5\times10^{-4}$} & \multirow{2}{*}{$9.9\times10^{-4}$} & \multirow{2}{*}{$0.017$} & \multirow{2}{*}{$0.026$} \\
  $R_{\mathrm{avg}} = 40.64\; \text{}, \alpha = 0.0015, \tau = 4.7052\;\text{}$ & & & & & & & \\\hline
%  Test data point with highest NRMSE in $\rho$ & \multirow{2}{*}{$3.7\times10^{-3}$} & \multirow{2}{*}{$1.0\times10^{-4}$} & \multirow{2}{*}{$4.3\times10^{-3}$} & \multirow{2}{*}{$3.1\times10^{-4}$} & \multirow{2}{*}{$8.9\times10^{-4}$} & \multirow{2}{*}{$0.059$} & \multirow{2}{*}{$0.027$} \\
%   $R=101.60\;\text{}, \alpha = 0.0000, \tau = 4.6052 \;\text{}$ & & & & & & & \\\hline
%   Test data point with highest NRMSE in $b$ & \multirow{2}{*}{$2.5\times10^{-2}$} & \multirow{2}{*}{$1.0\times10^{-5}$} & \multirow{2}{*}{$3.1\times10^{-2}$} & \multirow{2}{*}{$2.1\times10^{-4}$} & \multirow{2}{*}{$6.2 \times 10^{-3}$} & \multirow{2}{*}{$0.005$} & \multirow{2}{*}{$0.025$} \\
%   $R=76.17\;\text{}, \alpha = 0.0025, \tau = 4.4052 \;\text{}$ & & & & & & & \\\hline
     Test data point with unknown $R_{\mathrm{avg}}$: & \multirow{2}{*}{$6.7\times10^{-4}$} &\multirow{2}{*}{$1.9\times10^{-5}$} & \multirow{2}{*}{$4.2\times10^{-4}$} & \multirow{2}{*}{$4.3\times10^{-4}$} & \multirow{2}{*}{$6.7\times10^{-4}$} & \multirow{2}{*}{$0.018$} & \multirow{2}{*}{$0.028$} \\
  $R_{\mathrm{avg}} = 49.51\;\text{}, \alpha = 0.0020, \tau = 4.5052\;\text{}$ & & & & & & & \\\hline
   Test data point with unknown $\alpha$ & \multirow{2}{*}{$3.6\times10^{-3}$} &\multirow{2}{*}{$8.9\times10^{-5}$} & \multirow{2}{*}{$2.1\times10^{-3}$} & \multirow{2}{*}{$3.2\times10^{-4}$} & \multirow{2}{*}{$1.2\times10^{-3}$} & \multirow{2}{*}{$0.042$} & \multirow{2}{*}{$0.019$} \\
  $R_{\mathrm{avg}}=35.55\;\text{}, \alpha = 0.00125, \tau = 4.6052\;\text{}$ & & & & & & & \\\hline
   Test data point with unknown $\tau$ & \multirow{2}{*}{$2.2\times10^{-4}$} &\multirow{2}{*}{$5.4\times10^{-5}$} & \multirow{2}{*}{$2.7\times10^{-3}$} & \multirow{2}{*}{$4.7\times10^{-5}$} & \multirow{2}{*}{$2.9\times10^{-3}$} & \multirow{2}{*}{$0.008$} & \multirow{2}{*}{$0.023$} \\
 $R_{\mathrm{avg}} = 53.32\;\text{}, \alpha = 0.0015, \tau = 4.5552\;\text{}$  & & & & & & & \\\hline
   Test data point with unknown $R_{\mathrm{avg}}$, $\alpha$, $\tau$ & \multirow{2}{*}{$6.5\times10^{-4}$} &\multirow{2}{*}{$7.4\times10^{-5}$} & \multirow{2}{*}{$1.9\times10^{-3}$} & \multirow{2}{*}{$1.8\times10^{-4}$} & \multirow{2}{*}{$1.4\times10^{-3}$} & \multirow{2}{*}{$0.008$} & \multirow{2}{*}{$0.022$} \\
 $R_{\mathrm{avg}} = 49.51\;\text{}, \alpha = 0.00125, \tau = 4.5552\;\text{}$ & & & & & & & \\\hline 
   Test data point with unknown $R_{\mathrm{avg}}$, $\alpha$, $\tau$ & \multirow{2}{*}{$1.35\times10^{-4}$} &\multirow{2}{*}{$8.0\times10^{-5}$} & \multirow{2}{*}{$1.8\times10^{-3}$} & \multirow{2}{*}{$4.1\times10^{-4}$} & \multirow{2}{*}{$1.3\times10^{-3}$} & \multirow{2}{*}{$0.002$} & \multirow{2}{*}{$0.034$} \\
 $R_{\mathrm{avg}} = 30.46\;\text{}, \alpha = 0.00075, \tau = 4.6552\;\text{}$ & & & & & & & \\\hline 
\end{tabular}}
\caption{Errors in various post-processed quantities. Refer to eq.~\ref{eq:HF} and related discussion for interpretation of the various energetic terms. $R_{\mathrm{avg}}$ and $\tau$ values are in Bohr.}
\label{tab:Energy_Errors}
\end{table*}

We attribute the high accuracy of the model here to the accuracy in both dimensionality reduction and learning through NNs. The fact that during training, the model uses only about $120$ data points and that chemical accuracy requirements are met during prediction even for unseen input test cases, are particularly noteworthy and prove the effectiveness and generalizability of our model. Also, as pointed out earlier, in addition to being accurate, the proposed ML model is significantly more computationally efficient than DFT simulations. We also anticpate that the approach described here, i.e., obtaining complex electronic structure dependent quanties through post-processing of  ML predicted electronic fields, instead of predicting such quantities through ML directly, is going to find utility in computational studies related to the polarization and transport properties of low-dimensional systems.
\subsection{Interpretation of PCA modes} 
\begin{figure}[htbp]
    \centering
     \includegraphics[width=0.95\linewidth]{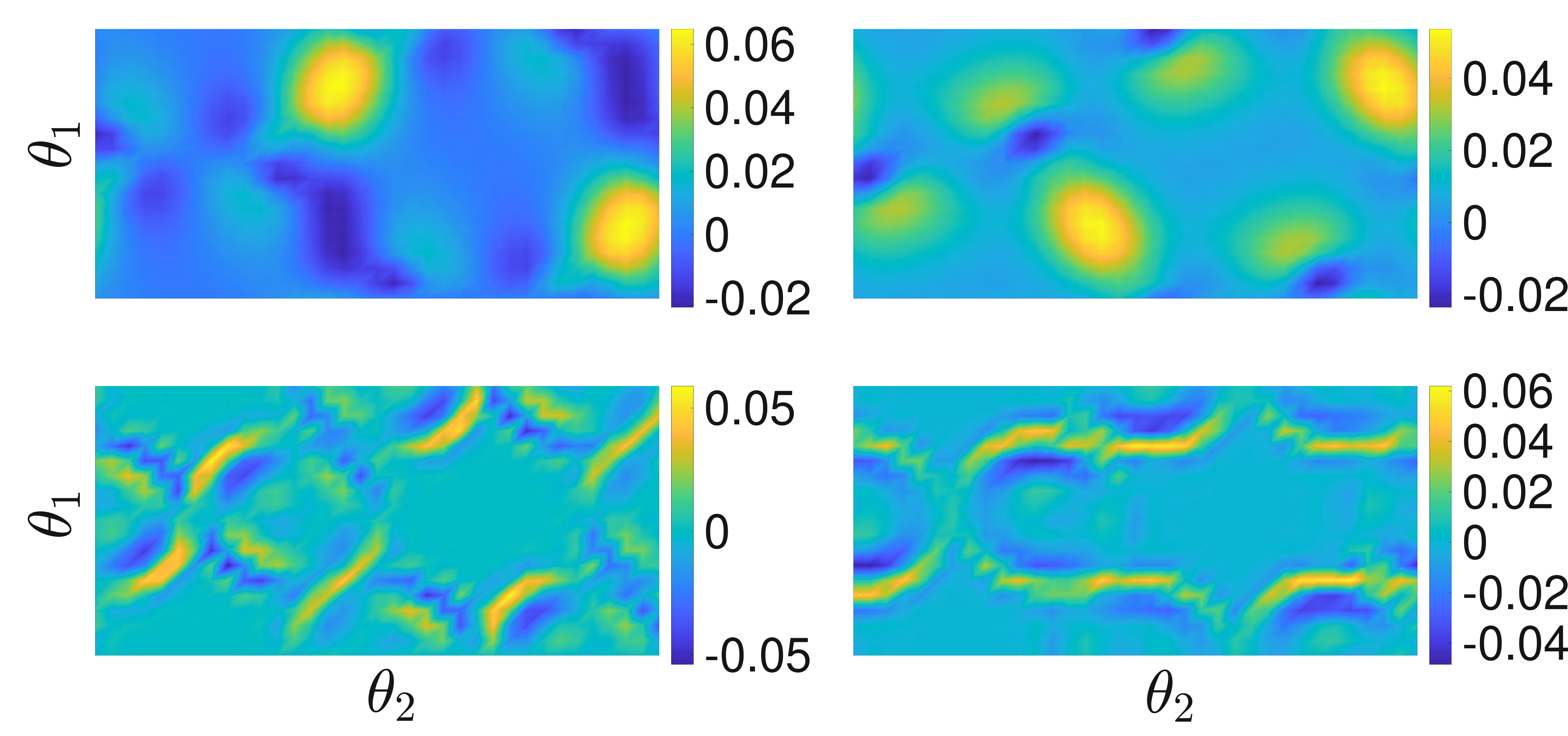}
    \caption{First two principal components for $\rho$ (\textit{top}) and $b$  (\textit{bottom}). {A slice of the PCA modes at the average radial coordinate of the atoms in the fundamental domain is shown}.}
    \label{fig:pca_modes}
\end{figure}

The number of PCs required in this problem is significantly less than the original dimensions of the electronic fields data ($\sim 60,\!000$), thus indicating that these quantities are mostly confined to subspaces of much lower--dimension. 
The presence of these hidden lower--dimensional features, and the significant reduction of dimensionality of the data through PCA, in turn, implies that just a few CoPCs have to be predicted as a function of the input parameters by the second step of the ML model. This helps account for the fact that such predictions can be made with relatively little training data, as discussed earlier. Remarkably, just the first couple of PCs appear sufficient to capture well over $90\%$ of the variance in both $\rho$ and $b$. Figure \ref{fig:pca_modes} shows these two PCA modes for each quantity visualized using helical coordinates, specifically in a $\theta_1-\theta_2$ plane located at the center of the simulation domain. As expected, the PCs of $\rho$ and $b$ capture the most significant aspects of the variations in these quantities, with the modes of $\rho$ reflecting changes in charge density along the carbon-carbon bonds, and those of $b$ capturing shifts in the nuclear positions. 

To elaborate on the above interpretations, we first recall that (see Section \ref{subsec:helical_coordinates}) in the absence of relaxation effects, an atom within the fundamental domain has the same values of $\theta_1$ and $\mathfrak{N}\,\theta_2$, regardless of the tube radius or the level of axial/torsional strain imposed (as before, $\mathfrak{N}$ denotes the cyclic symmetry group order). Consequently, for all values of the input parameters, the nuclei are expected to be located in the same relative positions in a $\theta_1-\theta_2$ planar plot, if atomic relaxation effects can be ignored. In the practice, upon imposition of strain, the nuclei re-adjust their positions to minimize the system energy during the structural relaxation procedure, leading to somewhat different values of the helical coordinates associated with their pseudocharge centers, than would be suggested by purely geometrical considerations. The PCA modes for $b$ illustrated in Figure \ref{fig:pca_modes} appear to be capturing this ``motion'' of the pseudocharge centers (i.e., nuclear coordinates) associated with the relaxation procedure. Furthermore, due to the changes in the nuclear positions, the carbon-carbon bond lengths change, and the PCA modes for $\rho$ appear to be capturing changes in the electron density along these bonds while they are stretched or compressed due to the imposed strains. Notably, these bonds are at angles with respect to the $\theta_1-\theta_2$ axes (see Figure \ref{fig:graphene_rolling} and top row of Figure \ref{fig:fields}), leading to the tilted appearance of the electron density lobes observable in Figure \ref{fig:pca_modes}. Finally, the presence of more wiggles in the plots for the PCA modes for $b$, as compared to those for $\rho$ can be explained by observing that at a discrete level, the latter is a smoother quantity. Specifically, the discretized $b$ field can have sharper local jumps since it is the sum of individual atom centered pseudocharges, while $\rho$ is more smeared out (also see top row of Figure \ref{fig:fields}). Indeed, this difference in relative degrees of smoothness at the discrete level probably contributes to the different number of PCA modes for these quantities needed to capture the same level of variance in the data (Figure \ref{fig:pca_var}).
%\subsection{Interpretation of the Principal Components}
%\red{Amartya Will Write **************}
\section{Conclusions and Future Directions}
This work proposes a machine learning model, which predicts electronic fields of quasi-one-dimensional materials under torsional and axial loads. We have demonstrated the utility of the technique by predicting the electron density and the nuclear pseudocharges for armchair carbon nanotubes as a function of the tube geometry and applied strains. The data generation process of the ML model uses a specialized symmetry adapted version of Kohn-Sham Density Functional Theory that is particularly well suited for the problem geometries and loading conditions considered here. The machine learning model has several salient features as we now summarize. \emph{First}, to populate the input space, quasi--random low--discrepancy sequences (Sobol sequence) are employed and DFT simulations are performed at these inputs to generate the data for training of the machine learning model. This strategy allows for obtaining an accurate machine learning model with a minimal number of data points ($123$ training simulations in this case). \emph{Second}, a two--step approach is taken to predict electronic fields. This involves dimensionality reduction of electronic fields followed by supervised learning in the reduced space. This two--step approach enables accurate prediction of high--dimensional electronic fields. The proposed ML model is remarkably accurate even for test cases with geometry and loading conditions that were not seen by the model during training. Moreover, the ML model is several orders of magnitude faster than the specialized, efficient Helical DFT technique used here for data generation. \emph{Third}, a  new technique based on a density--based clustering approach is developed to determine the atomic coordinates from the nuclear pseudocharges field predicted by the machine learning model. The atomic coordinates so obtained, are used to compute the non-local part of pseudopotentials that appear in the Kohn-Sham Hamiltonian, which would not have been possible otherwise. The electronic fields predicted by the machine learning model are postprocessed to obtain band structures, band gaps, total energies, and various energy components.  

We anticipate that machine learning models of the type developed here, will find use in computational investigations of strain engineering in low-dimensional systems and the multiscale modeling of the electromechanical response of such systems. One of the key advantages of the current ML model is its incorporation of symmetries commonly associated with quasi-one-dimensional systems, which makes it easier to explore the composition-structure space of such materials \citep{James_OS, Dumitrica_James_OMD}. Therefore, we anticipate that in conjunction with techniques for DFT calculations of large scale systems \citep{banerjee2018two, hu2015dgdft}, the current ML model and its extensions are likely to help in the exploration of novel phases of chiral matter \citep{qin2017superconductivity, aiello2022chirality} and compositionally complex nanotubes \citep{korde2022single}. Development of machine learning models that can capture atomic species specific features, as well as ones that can perform the post-processing steps associated with calculations of quantities such as energy components and band diagrams, serve as worthy directions for future research. 

\begin{acknowledgments}
SG acknowledges the financial support by NSF (MoMS) under grant number 1937983. SP and SG  acknowledge high--performance computing facilities:  (i)  Superior (at MTU) and (ii) Extreme Science and Engineering Discovery Environment (XSEDE) (allocation \#MSS200004). ASB acknowledges startup support from the Samueli School Of Engineering at UCLA, as well as funding from UCLA's Council on Research (COR) Faculty Research Grant. ASB and HMY would like to thank UCLA's Institute for Digital Research and Education (IDRE) for making available some of the computing resources used in this work. 
\end{acknowledgments}
\bibliography{main}% Produces the bibliography via BibTeX.
\appendix
%%%%%%%%%%%%%%%%%%%%%%%%%%%%%%%%%%%%%%%%%%%%%%%%%%%%%%%%%%%%%%%%%%%%%%%%%%%%%%%%%%%%%%%%%%%%%%%%%%%%%%%%%%
\section{Data Generation} \label{Appendix:dgc}
% \subsection{Data generation grid details} \label{Appendix:Data_gen}
\noindent \textbf{Data generation for Machine Learning}: 
We use the following bounds for the input space in order to generate the data: $R_{\mathrm{avg}} \in [20.32 \,, 101.60 \, ]$ Bohr, $\alpha \in [0,0.0025]$ and $\tau \in [4.4052,\; 4.8052]$ Bohr. This corresponds to choosing armchair CNTs with cyclic symmetry group orders between $16$ and $80$, i.e., with radii in the experimentally relevant $1$ to $5$ nanometer range. The increments in $\alpha$ and $\tau$ are $0.0005$ and $0.1$ Bohr, respectively. The maximum applied torsional strain of about $3.86$ degrees/nm --- close to the regime in which torsional instabilities may start to appear \citep{Dumitrica_James_OMD}, and the maximum axial strain considered here is about $4.3 \%$. Helical DFT simulations were performed in the input space following Sobol sequencing. Note that, the Sobol sequence would not always generate a sample point that is feasible for simulations, given the discrete nature of the nanotube radius. For such cases, we have carried out simulations at the nearest feasible value of $R_{\mathrm{avg}}$ and strain parameters.
%As mentioned earlier, simulation data was obtained for 164 data points in the input variables space such that $R_{\mathrm{avg}} \in [20\, \mathrm{Bohr}, \; 102 \,\mathrm{Bohr}] $, $\alpha \in [0,3.69\degree/\mathrm{nm}]$ and $\tau \in [4.4052 \, \text{Bohr},4.8052 \, \text{Bohr}]$. Simulations were performed in the input space following the Sobol sequencing. Note that, the Sobol sequence would not always give the sample point that is feasible for simulations given the discrete nature of $R_{\text{avg}}$ and possible increments in the loading. For such cases we have simulated at the nearest location in the input space. 
% Our simulation data points have the following paramter values, $R_{\mathrm{avg}} = $ 20.32,  \; 25.39, \;  27.93, \;  35.55, \;  40.64, \;  46.97, \;  53.32, 60.96, \;  66.02,  \; 71.09, \;  76.17, \;  81.28, \; 101.60 (Bohr); $\alpha =$ 0,\; 0.0005,\; 0.001,\; 0.0015,\; 0.002,\; 0.0025; $\tau =$ 4.4052,\; 4.5052,\; 4.6052,\; 4.7052,\;  4.8052. 

\begin{figure}[htbp]
    \centering
    \includegraphics[width=0.49\linewidth]{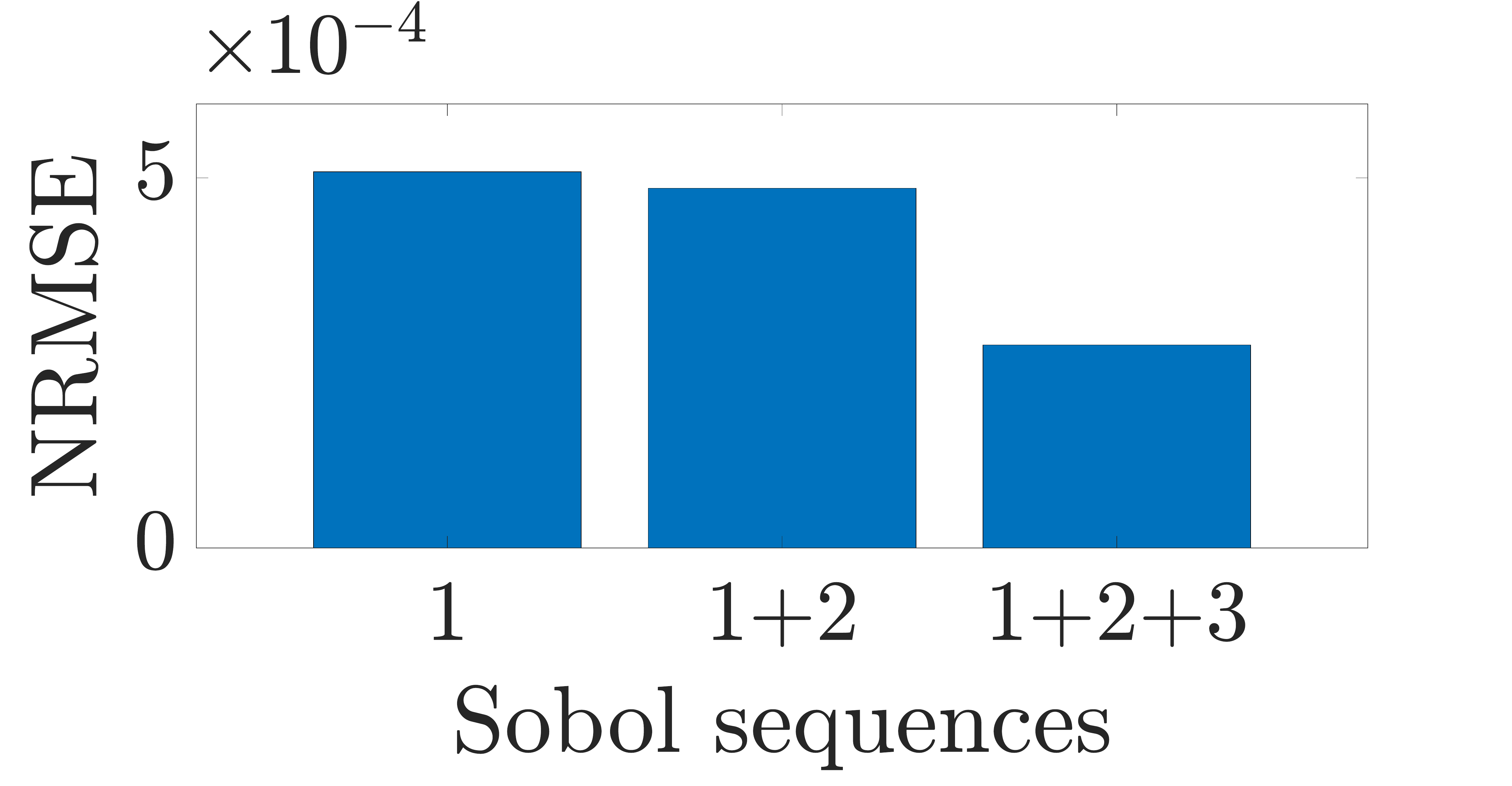}
    \includegraphics[width=0.49\linewidth]{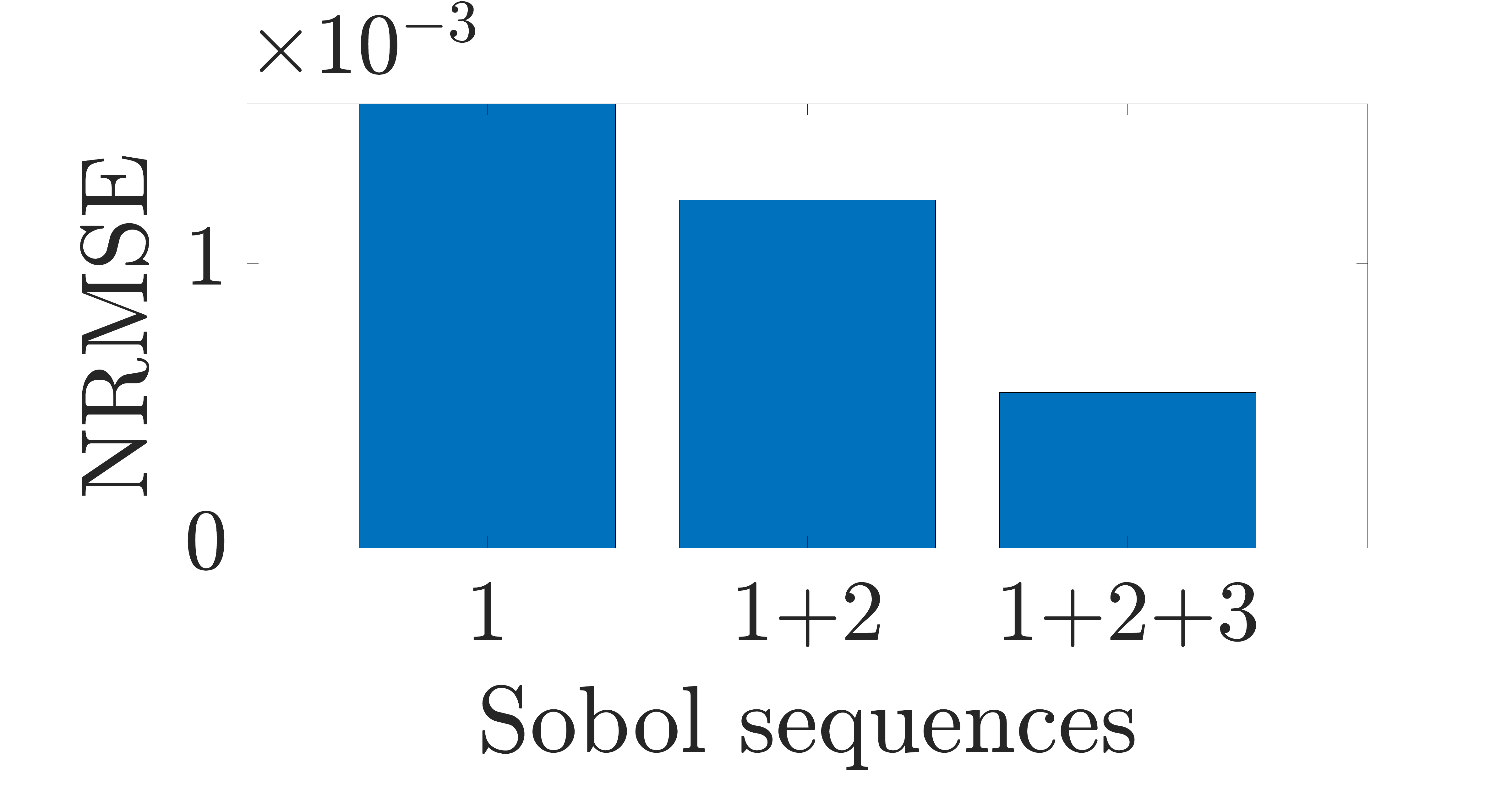}
    \caption{Mean of NRMSE for test data when the machine learning model is trained using Sobol sets.  (\emph{Left}) Error bars for $\rho$, (\emph{Right}) Error bars for $b$.}
    \label{fig:error_sobolpca}
\end{figure}

To achieve the desired accuracy in the prediction of the electronic fields with the minimum number of DFT simulations we start with a set of points guided by the Sobol sequence. Subsequently, we add simulations in smaller sets (referred to as Sobol sets here) to the training data, till we attain the desired accuracy in the prediction of electronic fields. Our first set consists of 85 simulations followed by 48 and 31 simulations in the second and third sets, respectively. As mentioned in Section \ref{subsec:Data_gen_ML} these three sets of simulations successively refine the input space. Fig. \ref{fig:error_sobolpca} shows NRMSE for test data obtained when the ML model was trained using these three Sobol sets cumulatively. The first bar denotes test set NRMSE when only set I (85 data points) was used; the second bar denotes test set NRMSE when sets 1 and 2 (85+48 data points) were used; the third bar denotes test set NRMSE when sets 1,2 and 3 (85+48+31 data points) were used. Note that for each of these cases, $15$\% of the data available to the ML model was used for testing.
% This forms a total of 390 simulation points. From this 390 points we simulated (DFT) at only 164 points following Sobol sequencing as mentioned in the main manuscript. 
% We first formed a grid along $R_{\mathrm{avg}}$, $\alpha$ and $\tau$ direction. Note that, the Sobol sequence would not always give the sample point lying on the chosen grid. In that case we have simulated at the nearest location in the input space grid.  Along $R_{\mathrm{avg}}$ direction we took $R_{\mathrm{avg}} = $ 20.32,  \; 25.39, \;  27.93, \;  35.55, \;  40.64, \;  46.97, \;  53.32, 60.96, \;  66.02,  \; 71.09, \;  76.17, \;  81.28, \; 101.60 (Bohr), along $\alpha$ direction we took $\alpha =$ 0,\; 0.0005,\; 0.001,\; 0.0015,\; 0.002,\; 0.0025 and along $\tau$ direction we took $\tau =$ 4.4052,\; 4.5052,\; 4.6052,\; 4.7052,\;  4.8052. Note that in the results section of the main paper, wherever it is mentioned that the input parameter is unseen, it means that the value of the input parameter is beyond the values of the input parameters mentioned above. This grid results in a total 390 grid points. From this 390 grid points we simulated (DFT) at 164 grid points following Sobol sequencing as mentioned in the main manuscript. 
\section{Training of Neural Networks}
\label{Appendix:DNN_training}
\begin{figure}[htbp]
    \centering
    \includegraphics[width=0.99\linewidth]{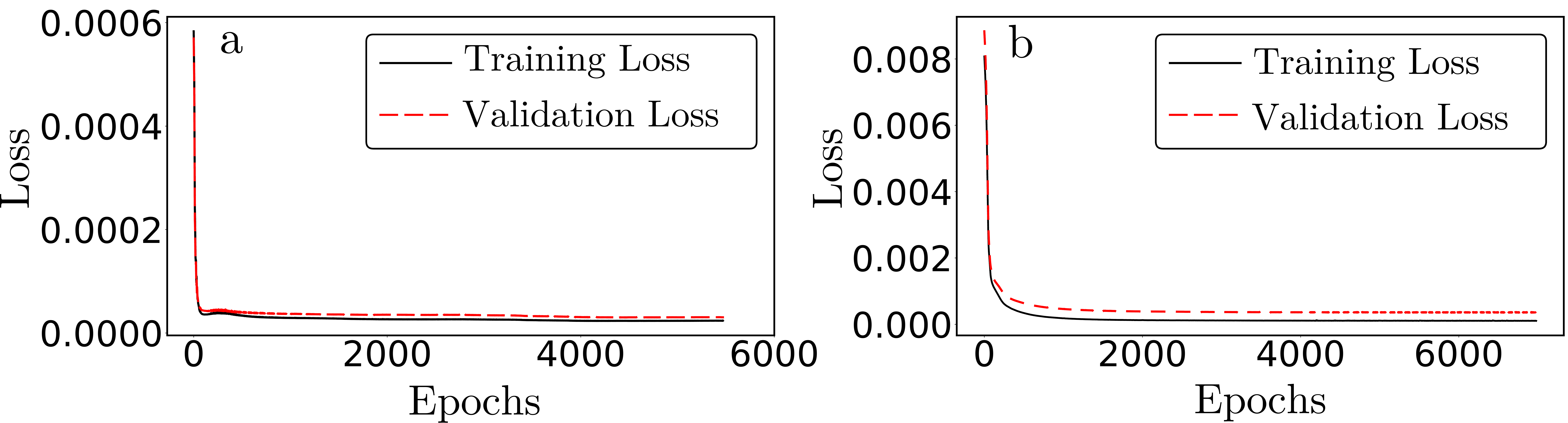}
    \caption{(a) Learning curve for $\mathcal{N}_1$, (b) Learning curve for $\mathcal{N}_2$.}
    % Loss plotted here is the mean squared error over the true and predicted electronic fields.}
    \label{fig:learning_curves}
\end{figure}

The Learning curves for the neural networks $\mathcal{N}_1$ (for $\rho$) and $\mathcal{N}_2$ (for $b$) are presented in Fig. \ref{fig:learning_curves}. The loss function used to train the neural networks is computed on CoPCs and is given in Eq. \ref{eq:loss_with_reg} below.
%\newpage
\noindent \textbf{Hyperparameter Optimization and Regularization} \\
The proposed machine learning model contains various parameters associated with the NNs that can affect the model's overall accuracy and performance. In particular, so-called hyperparameters associated with controlling the learning process need to be tuned. Important hyperparameters for our model include the architectures for NNs, the activation function, the learning rate, and the number of iterations. We discuss each of these below.

\begin{figure}[htbp]
    \centering
    \includegraphics[width=0.75\linewidth]{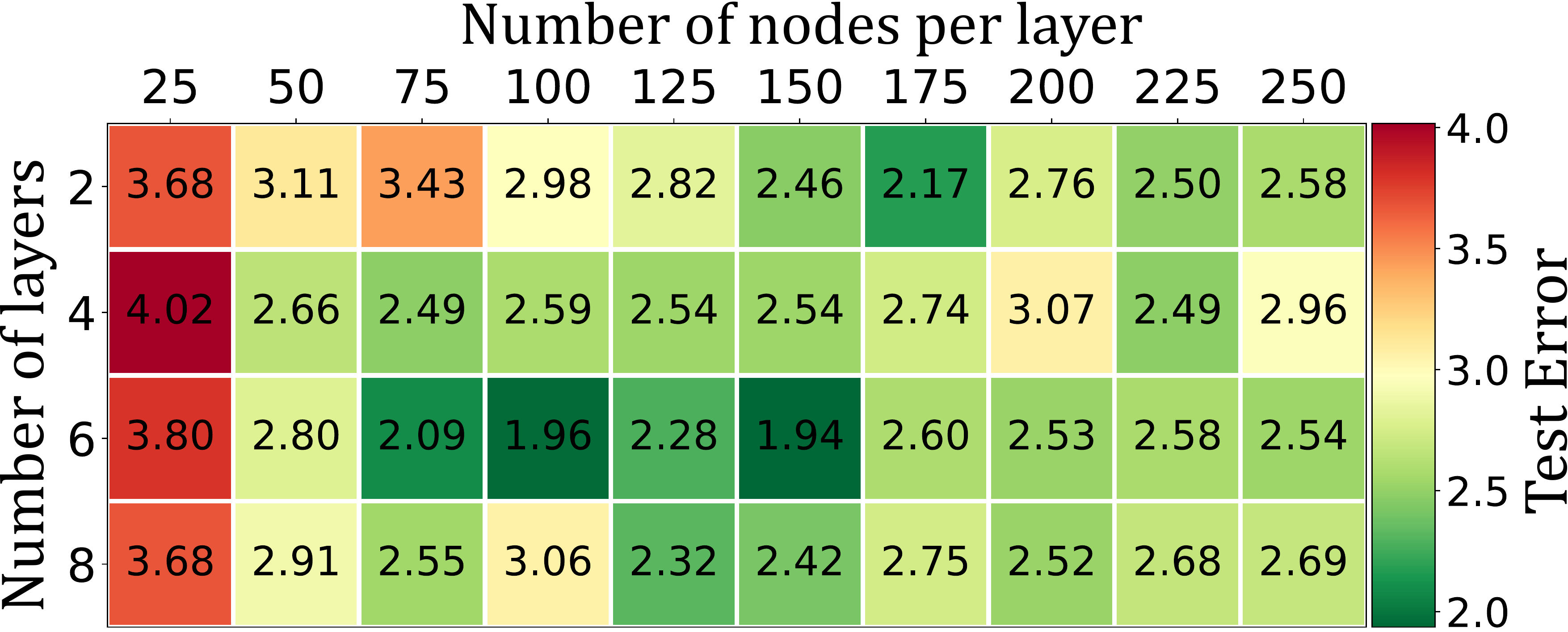}
    \includegraphics[width=0.75\linewidth]{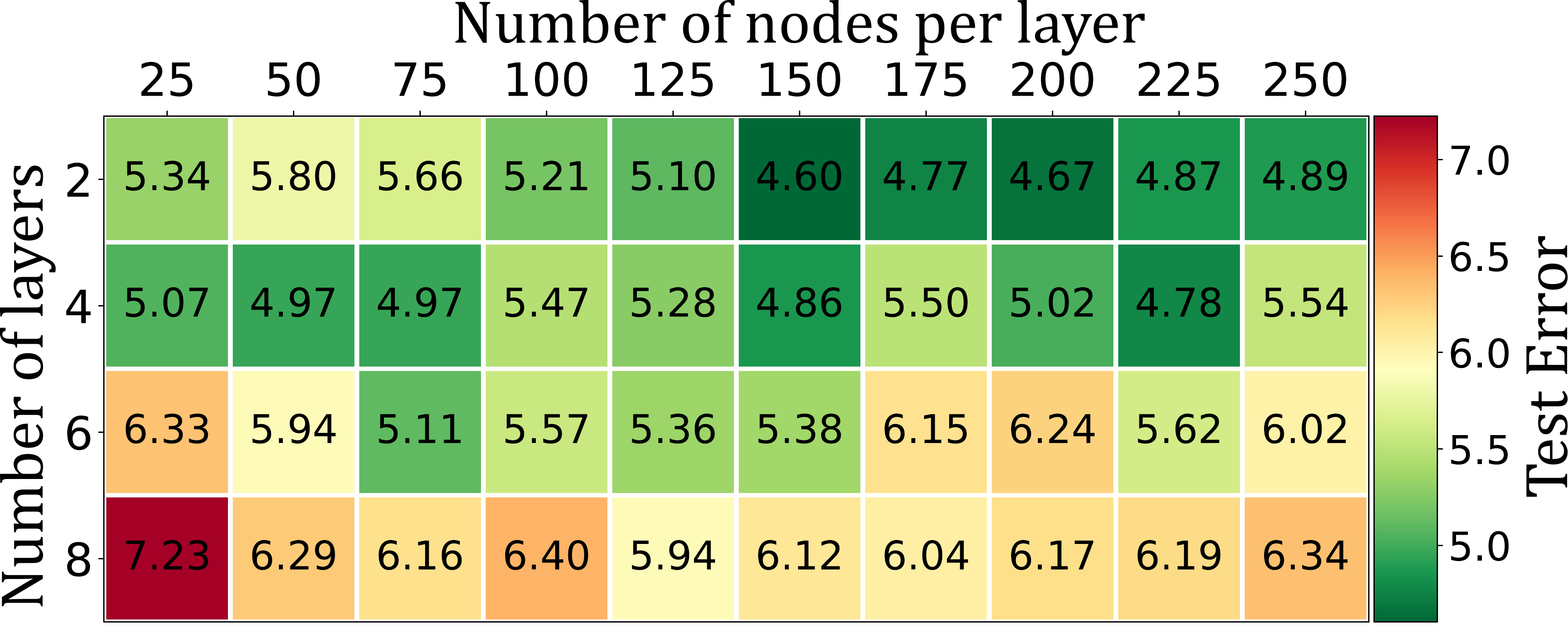}
    \caption{(\emph{Top}) Test error ($\times 10^{-5}$) for various architectures of $\mathcal{N}_1$ (trained for $\rho$), (\emph{Bottom}) Test error ($\times 10^{-4}$) for various architectures of $\mathcal{N}_2$ (trained for $b$.)} 
    % \red{\textbf{Can this figure be made larger and/or in higher quality ? It's not looking so great in the current form and the numbers are barely readable.}}}
    \label{fig:arch_selection}
\end{figure}

\noindent \textbf{Architecture:} The predictive capability of a NN and the accuracy obtained in the prediction depends on the number of hidden layers and the number of neurons in the hidden layers. We optimized the number of hidden layers and number of neurons per layer using the grid search method \cite{liashchynskyi2019grid}. Fig. \ref{fig:arch_selection} shows the test error for $\mathcal{N}_1$ and $\mathcal{N}_2$ trained for varying number of layers and varying number of nodes per layer. For $\mathcal{N}_1$, six layers of $150$ neurons each yielded the least test error, and for $\mathcal{N}_2$, two layers of $150$ neurons each yielded the least test error.  

\noindent \textbf{Activation Function:} We used Rectified  Linear  Unit  (ReLU) as an activation function for both neural networks $\mathcal{N}_1$ and $\mathcal{N}_2$.  This choice avoids problems of vanishing or exploding gradients encountered by other common activation functions like Sigmoid, Tanh \cite{glorot2011deep, glorot2010understanding}.

\noindent \textbf{Learning Rate:} The learning rate was set to $0.001$ as suggested in \cite{kingma2014adam}. Other parameters pertinent to the Adam optimizer were set at suggested values \cite{kingma2014adam}  based on good results for other machine learning problems ($\beta_1 = 0.9$, $\beta_2 = 0.999$, $\epsilon = 10^{-8}$).

\noindent \textbf{Number of Iterations:} In order to avoid overfitting and help ensure good generalization performance of the ML model, we used early stopping \cite{Prechelt2012,sarle1996stopped}. We stopped the training when the validation loss does not improve over a specific number of \emph{patience epochs}. We employed \emph{patience} epochs of $1000$, and the maximum number of epochs was set to $200000$.  
% Early stopping is a technique to stop the training process before completing the maximum epochs by monitoring the validation loss. An increase in the validation loss while the training loss decreases is probable when the model overfits the data. Early stopping avoids overfitting by stopping the training when the validation loss does not improve over a specific number of epochs, called \emph{patience} epochs. In our work, we employed early stopping with \emph{patience} epochs of 1000, and the maximum number of epochs was set at 200000.  The use of early stopping ensured the generalization performance of our model and avoided overfitting. 

\noindent \textbf{Elastic net regularization:} Along with early stopping, we used elastic net regularization to avoid overfitting. This technique is a combination of $\mathcal{L}_{1}$ and $\mathcal{L}_{2}$ regularization methods \cite{zou2005regularization}, and overcomes the individual drawbacks of each. The loss function including $\mathcal{L}_{1}$ and $\mathcal{L}_{2}$ regularization can be written as:
\begin{eqnarray}\label{eq:loss_with_reg}
    {
    \tilde{\mathcal{J}}(y, \mathcal{N}(x, \bar{\mathbf{w}}))}  & {=  \mathcal{J}(y, \mathcal{N}(x, \bar{\mathbf{w}})) + \lambda_1||\bar{\mathbf{w}}||_1}
    { + \lambda_2||\bar{\mathbf{w}}||_2}\,,
    {\qquad \lambda_1, \lambda_2 \in \mathbb{R}}\,.
\end{eqnarray}
Here, $\mathcal{J}(y, \mathcal{N}(x, \bar{\mathbf{w}}))$ is the mean squared error over the true outputs $y$ and the neural network ($\mathcal{N}$) predicted outputs $y' = \mathcal{N}(x, \bar{\mathbf{w}})$. Furthermore,  $\bar{\mathbf{w}}$ are the weights and biases of $\mathcal{N}$, $x$ is the input, and ${ ||\bar{\mathbf{w}}||_{1}}$ and  
${||\bar{\mathbf{w}}||_{2}}$ are the ${\mathcal{L}_{1}}$ and ${\mathcal{L}_{2}}$ regularization terms, respectively. We have used $\lambda_1 = \lambda_2 = 10^{-5}$ for both $\mathcal{N}_1$ and $\mathcal{N}_2$. %\red{Check the equation !}

\section{Comparison of the Clustering and Neural Network Approaches to Obtaining the Nuclear Coordinates} \label{app:comparison_nn_dbscan}
\begin{figure}[h!]
    \centering
    \includegraphics[width=0.75\linewidth]{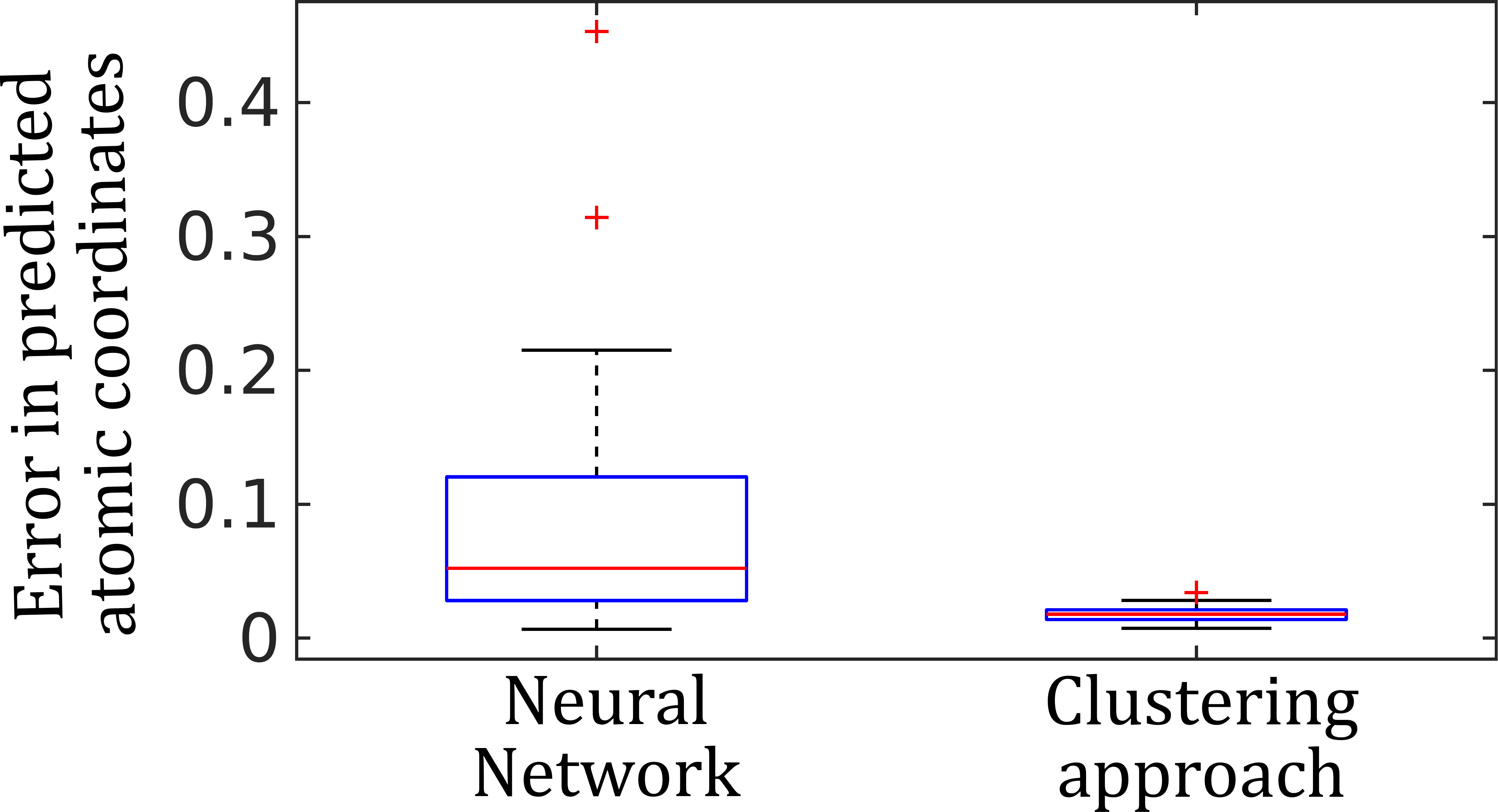}
    \caption{Error in predicted atomic coordinates (in Bohr), i.e., the distance between true and predicted nucleus positions, using a neural network and the DBSCAN based clustering approach.} 
    \label{fig:atomic_ccordinates_error}
\end{figure}

We compare our clustering based approach to determine atomic coordinates with a neural network that was trained to predict the atomic coordinates directly from the inputs 
% to the neural network models for $\rho$ and $b$, i.e., 
: $R_{\mathrm{avg}}, \alpha$ and $\tau$. We found that the error (distance between actual atomic coordinates and predicted atomic coordinates) was significantly higher in the case of the neural network model than the DBSCAN based approach proposed here. The errors in atomic coordinates using the neural network approach and our clustering approach are shown in Fig \ref{fig:atomic_ccordinates_error}. The superior performance of the clustering based approach is likely related to the ability of the method to make use of the specific structure of the $b$ field (i.e., it is the superposition of a set of non-overlapping, atom-centered, spherically symmetric charge distributions), as opposed to the neural network model which does not incorporate such information.

\newpage
{\section{Comparison of our approach with a direct Neural Network based prediction of the band gap location} \label{app:bandloc_nn_ourapproach}}
{It is well known \citep{yu2022density, ding2002analytical} that the location of the band gap in an armchair carbon nanotube shifts in  the $(\eta,\nu)$ space, as the tube is distorted. For small twists in particular, the bandgap continues to be a direct one and remains at $\eta = \frac{1}{3}$, while transitioning to different values of the cyclic symmetry index $\nu$. We show in Table \ref{table:bandgap_loc} below that for the same amount of training data, this behavior is correctly captured by our method of predicting band structure dependent quantities (i.e., carrying out a low-overhead postprocessing step on the ML predicted smooth electronic fields), in contrast to a direct neural network based prediction.}

\begin{table}[h!]
\begin{tabular}{|c|ccc|ccc|}
\hline
\multirow{2}{*}{Case} & \multicolumn{3}{c|}{$\nu$ for VBM} & \multicolumn{3}{c|}{$\nu$ for CBM} \\ \cline{2-7} 
 & \multicolumn{1}{c|}{DFT} & \multicolumn{1}{c|}{Our Approach} & NN & \multicolumn{1}{c|}{DFT} & \multicolumn{1}{c|}{Our Approach} & NN \\ \hline
\begin{tabular}[c]{@{}c@{}}Random test data point\\ $R_{\mathrm{avg}} = 40.64\; \text{}, \alpha = 0.0015, \tau = 4.7052\;\text{}$\end{tabular} & \multicolumn{1}{c|}{2} & \multicolumn{1}{c|}{2} & 1.76 & \multicolumn{1}{c|}{2} & \multicolumn{1}{c|}{2} & 1.82 \\ \hline
\begin{tabular}[c]{@{}c@{}}Test data point with unknown $R_{\mathrm{avg}}$ \\ $R_{\mathrm{avg}} = 49.51\;\text{}, \alpha = 0.0020, \tau = 4.5052\;\text{}$\end{tabular} & \multicolumn{1}{c|}{3} & \multicolumn{1}{c|}{3} & 3.35 & \multicolumn{1}{c|}{3} & \multicolumn{1}{c|}{3} & 3.15 \\ \hline
\begin{tabular}[c]{@{}c@{}}Test data point with unknown $\alpha$\\ $R_{\mathrm{avg}}=35.55\;\text{}, \alpha = 0.00125, \tau = 4.6052\;\text{}$\end{tabular} & \multicolumn{1}{c|}{1} & \multicolumn{1}{c|}{1} & 1.06 & \multicolumn{1}{c|}{1} & \multicolumn{1}{c|}{1} & 1.12 \\ \hline
\begin{tabular}[c]{@{}c@{}}Test data point with unknown $\tau$ \\ $R_{\mathrm{avg}} = 53.32\;\text{}, \alpha = 0.0015, \tau = 4.5552\;\text{}$\end{tabular} & \multicolumn{1}{c|}{3} & \multicolumn{1}{c|}{3} & 2.91 & \multicolumn{1}{c|}{3} & \multicolumn{1}{c|}{3} & 2.86 \\ \hline
\begin{tabular}[c]{@{}c@{}}Test data point with unknown $R_{\mathrm{avg}}$, $\alpha$, $\tau$\\  $R_{\mathrm{avg}} = 49.51\;\text{}, \alpha = 0.00125, \tau = 4.5552\;\text{}$\end{tabular} & \multicolumn{1}{c|}{2} & \multicolumn{1}{c|}{2} & 2.06 & \multicolumn{1}{c|}{2} & \multicolumn{1}{c|}{2} & 2.06 \\ \hline
\begin{tabular}[c]{@{}c@{}}Test data point with unknown $R_{\mathrm{avg}}$, $\alpha$, $\tau$\\ $R_{\mathrm{avg}} = 30.46\;\text{}, \alpha = 0.00075, \tau = 4.6552\;\text{}$\end{tabular} & \multicolumn{1}{c|}{0} & \multicolumn{1}{c|}{0} & 0.4 & \multicolumn{1}{c|}{0} & \multicolumn{1}{c|}{0} & 0.36 \\ \hline
\end{tabular}
\caption{{Comparison of our approach (ML model followed by postprocessing) with direct neural network (NN) based prediction of cyclic symmetry index ($\nu$) associated with band-gap location. VBM denotes the valence band maximum and CBM denotes the conduction band minimum. DFT denotes reference first principles results calculated using Helical DFT. $R_{\mathrm{avg}}$ and $\tau$ values are in Bohr.}}
\label{table:bandgap_loc}
\end{table}

\end{document}